\documentclass{article}
\usepackage{latexsym}
\usepackage{amssymb}
\usepackage{amsmath}
\usepackage[numbers,sort&compress]{natbib}
\usepackage{color}
\usepackage{amsfonts,amssymb}
\usepackage{mathrsfs}
\usepackage{dsfont}
\usepackage{graphicx}

\newcommand{\csch}{\mathrm{csch}}

\def\cdot{{\scriptstyle\,\bullet\,}}

\begin{document}
\title{\bf Two integrable differential-difference equations derived from NLS-type equation}
\author{Zong-Wei Xu$^\dag$,Guo-Fu Yu$^\dag$ and Yik-Man Chiang$^\ddag$,\\
$^\dag$Department of Mathematics, Shanghai Jiao Tong University, \\
Shanghai 200240, P.R.\ China\\
$^\ddag$ Department of Mathematics, Hong Kong University of Science
and
Technology,\\ Clear Water Bay, Kowloon, Hong Kong,\ P. R. China
}
\date{}
\maketitle

\begin{abstract}
Two integrable differential-difference equations are derived from a
(2+1)-dimensional modified Heisenberg ferromagnetic equation and a resonant nonlinear
Schr\"oinger equation respectively.
Multi-soliton solutions of the resulted semi-discrete systems are
given through Hirota's bilinear method. Elastic and inelastic interaction behavior
between two solitons are studied through the asymptotic
analysis. Dynamics of two-soliton solutions are shown with graphs.
\end{abstract}

{\bf keywords:} modified Heisenberg ferromagnetic system;
Resonant nonlinear Schr\"oinger equation;
Integrable discretization; Soliton interactions.

\section{Introduction}
Recently, integrable discretizations of integrable equations have
been of considerable and current interest in soliton theory.
As Suris mentioned \cite{suris}, various approaches to the
problem of integrable discretization are currently developed, among which the
Hirota's bilinear method is very powerful and effective. Discrete analogues of
almost all interesting soliton equations, the KdV, the Toda chain, the sine-Gordon, etc.,
can be obtained by the Hirota method. The purpose of this paper is to consider
integrable discrete analogues of two nonlinear Schr\"odinger (NLS)-type system by Hirota method.

The one-dimensional classical continuum Heisenberg models with
different magnetic interactions have been settled as one of the
interesting and attractive classes of nonlinear dynamical equations
exhibiting the complete integrability on many occasions. As is well
known, Heisenberg first proposed in 1928 the following discrete
(isotropic) Heisenberg ferromagnetic (DHF)  spin chain \cite{HS}
\begin{align}\label{DHF}
\dot{\mathbf{S}}_n=\mathbf{S_n}\times(\mathbf{S_{n+1}}+\mathbf{S_{n-1}}),
\end{align}
where $\mathbf{S_n}=(\mathrm{s}_1^n,\mathrm{s}_2^n,\mathrm{s}_3^n)\in \mathbf{R}^3$ with
$|\mathbf{S_n}|=1$ and the overdot represents the time derivative
with respect to $t$. The DHF chain plays an important role in the
theory of magnetism.

A performance of the standard continuous limit procedure leads DHF
model \eqref{DHF} to the integrable Heisenberg ferromagnetic model
\begin{align}\label{HF}
\mathbf{S}_t=\mathbf{S}\times \mathbf{S}_{xx},
\end{align}
which is an important equation in condensed matter physics
\cite{cHS}. NLS-type equations are
extensively used to describe nonlinear water waves in fluids,
ion-acoustic waves in plasmas, nonlinear envelope pulses in the
fibers. It is known that HF is gauge equivalent to NLS equation and
DHF is gauge equivalent to a kind of discrete NLS-like equation
\cite{Ding}.

Higher dimensional nonlinear evolution equations are proposed to
describe certain nonlinear phenomena. Due to the dependence on the
additional spatial variables in higher dimensional systems, richer
solution structure might appear, such as dromions, lumps, breathers
and loop solitons.

An extension of Eq. \eqref{HF} is the $(2 + 1)$-dimensional
integrable modified HF system \cite{Anco,Duan07,Thomas, Victor}, as
follows,
\begin{subequations}\label{HF1}
\begin{eqnarray}
&& u_t+u_{xy}+uw=0,\label{non1}\\
&& v_{t}-v_{xy}-vw=0,\label{non2}\\
&& w_x+(uv)_{y}=0,\label{non3}
\end{eqnarray}
\end{subequations}
which is associated with the $(2+1)$-dimensional $\mathrm{NLS}$
equation
\begin{align}
\textrm{i}q_{\tau}+q_{\xi\tau}-2q\int (|q|^2q)_{\eta}d\xi=0,
\end{align}
where $\eta$ is a spatial variable $y$. System \eqref{HF1} can also
be used to model the biological pattern formation in
reaction-diffusion process \cite{Victor,Koch}. In Ref.
\cite{Duan07}, system \eqref{HF1} was investigated through the
prolongation structure and Lax representation. In Refs.
\cite{Thomas,Victor}, integrable property of system \eqref{HF1} was
studied through the Painlev\'e analysis, and some localized coherent
and periodic solutions were given by means of the multi-linear
variable separation approach. Multi-soliton solutions of system
\eqref{HF1} was derived in \cite{Tian2014} by means of the Hirota
bilinear method, and the double Wronskian solutions was given
therein. Similar as the counterpart between DHF \eqref{DHF} and continuous HF
\eqref{HF}, it is natural to consider discrete version of the
$(2 + 1)$-dimensional modified HF system \eqref{HF1}.

A new integrable version of the
nonlinear Schr\"odinger (NLS) equation, called the ${\it resonant}$
NLS (RNLS) equation,
\begin{align}
\rm{i}U_t+U_{xx}+\frac{\alpha}{4}|U|^2U=\beta\frac{1}{|U|}|U|_{xx}U,\label{RS}
\end{align}
was recently proposed \cite{PL02}
to describle low-dimensional gravity (the Jackiw-Teitelboim model)
and  response of a medium to the action of a quasimonochromatic wave.
Here $\alpha$ is a nonlinear coefficient, and $\beta$ denotes the
strength of electrostriction pressure or diffraction. The term
$|U|_{xx}/|U|$ on the righthand side of Eq. \eqref{RS} is so called
the "quantum potential". Moreover, Eq.
\eqref{RS} can model propagation of one-dimensional long
magnetoacoustic waves in a cold collisionless plasma subject to a
transverse magnetic field \cite{LPR07}. Note that when $\beta < 1$
Eq. \eqref{RS} is reduced to NLS
\begin{align}
i\Phi_{\xi}+\Phi_{\tau\tau}+\sigma |\Phi|^2\Phi=0.\label{NLS}
\end{align}
When $\beta
> 1$, it is not reducible to the usual NLS equation
but to a reaction-diffusion (RD) system \cite{LPR07,LP07,PLR08}.

The Lax pair in the $2\times 2$ matrix form of Eq.\eqref{RS} was given in \cite{PLANZ02}.
Based on the RD version of RNLS,
multi-soliton solutions were derived via the B\"acklund-Darboux
transformations in \cite{LPR07} and via the Hirota method in
\cite{LP07} respectively.  In \cite{PLR08}, integrable extension of
Eq.\eqref{RS} has been suggested and soliton solutions of the
resonant NLS case under the reduction condition have been presented
via the Hirota method. Additionally, the method of binary Bell
polynomials was used to study two-soliton solution and integrable
properties of Eq.\eqref{RS} in \cite{TB13}.

Since NLS equation has integrable discrete versions and RNLS is an
intermediate equation between the focusing and defocusing NLS
equation, it is natural to study the discrete analogue of RNLS. This paper is to investigate the integrable
discretization of RNLS equation \eqref{RS} in the case of $\beta>1$.

The paper is structured as follows. In section 2, we present
multi-soliton solutions and discrete version of $(2+1)-$dimensional modified HF equation.
In section 3, we study RNLS equation, including multi-soliton solutions, semi-discrete analogue
and dynamic properties of solutions.
Finally, a short conclusion is given in section 4.

\section{(2+1)-dimensional modified HF equation}

\subsection{$N$-soliton solution to modified HF system}
Through the variable transformation
\begin{align}
u=\frac{G}{F},\quad v=\frac{H}{F},\quad w=2(\ln(F))_{xy},
\end{align}
Eq.\eqref{HF1} transforms into the following bilinear form
\begin{subequations}
\label{2HF}
\begin{align}
& (D_t+D_{x}D_y) G\cdot F=0,\label{hfb1}\\
& (D_t-D_{x}D_y) H\cdot F=0,\label{hfb2}\\
& D_x^2F\cdot F+GH=0.\label{hfb3}
\end{align}
\end{subequations}
The Hirota bilinear differential operator $D^m_xD_t^k$ is  defined
by \cite{AJG}\\
\begin{align*}
&D_x^mD_y^n a\cdot b\equiv \left (\frac \partial {\partial
x}-\frac
\partial {\partial x'}\right )^m\left (\frac \partial {\partial
y}-\frac
\partial {\partial y'}\right )^na(x,y)b(x',y')|_{x'=x,y'=y},\nonumber\\
& m,n=0,1,2,\cdots.
\end{align*}
One-, two- and three-soliton solutions of bilinear equation set \eqref{2HF} are given in
\cite{Tian2014}. Note that the bilinear equations \eqref{2HF} are in Schr\"odinger type,
here we present a compact form of multi-soliton
solutions to the system \eqref{2HF}. The two-soliton is given as
\begin{align}
&F=1+a(1,1^*)\exp(\eta_1+\xi_1)+a(1,2^*)\exp(\eta_1+\xi_2)\nonumber\\
&\qquad+a(2,1^*)\exp(\eta_2+\xi_1)
+a(2,2^*)\exp(\eta_2+\xi_2)\nonumber\\
&\qquad+a(1,2,1^*,2^*)\exp(\eta_1+\eta_2+\xi_1+\xi_2),\\
&
G=\exp(\eta_1)+\exp(\eta_2)+a(1,2,1^*)\exp(\eta_1+\eta_2+\xi_1)\nonumber\\
&\qquad+a(1,2,2^*)\exp(\eta_1+\eta_2+\xi_2),\\
&
H=\exp(\xi_1)+\exp(\xi_2)+a(1,1^*,2^*)\exp(\eta_1+\xi_1+\xi_2)\nonumber\\
&\qquad+a(2,1^*,2^*)\exp(\eta_2+\xi_1+\xi_2).
\end{align}
with
\begin{align}
& \eta_i=k_ix+l_iy+\omega_i t+\eta_i^0,\quad w_i=-k_il_i,\\
& \xi_i=p_ix+q_iy+\Omega_it+\xi_i^0,\quad \Omega_i=p_iq_i,
\end{align}
and the coefficients are defined by
\begin{align}
& a(i,j)=-2(k_i-k_j)^2,\label{ce1}\\
& a(i,j^*)=-\frac{1}{2(k_i+p_j)^2},\\
& a(i^*,j^*)=-2(p_i-p_j)^2,\\
&a(i_1,i_2,\cdots,i_n)=\prod_{1\leq l<k\leq n}a(i_l,i_k).\label{pfa}
\end{align}

Generally, exact $N$-soliton solution to Eq. \eqref{2HF}
is expressed in the following form
\begin{eqnarray}
&& F=\sum_{\mu=0,1}^{(e)}\exp\left[\sum_{j=1}^{N}\mu_j\eta_j
+\sum_{j=N+1}^{2N}\mu_j\xi_{j-N}+\sum_{1\leq
i<j}^{2N}\mu_i\mu_jA_{ij} \right],\label{csolu1}\\
&& G=\sum_{\nu=0,1}^{(o)}\exp\left[\sum_{j=1}^{N}\nu_j\eta_j+
\sum_{j=N+1}^{2N}\nu_j\xi_{j-N}+\sum_{1\leq
i<j}^{2N}\nu_i\nu_jA_{ij}
\right],\label{csolu2}\\
&&
H=\sum_{\lambda=0,1}^{(o)}\exp\left[\sum_{j=1}^{N}\lambda_j\eta_j+
\sum_{j=N+1}^{2N}\lambda_j\xi_{j-N}+\sum_{1\leq
i<j}^{2N}\lambda_i\lambda_jA_{ij} \right],\label{csolu3}
\end{eqnarray}
with
\begin{eqnarray}
&& \eta_j=k_j x+l_j y+\omega_jt,\quad \omega_j=-k_jl_j,\quad
j=1,2,\cdots,N,\\
&& \xi_j=p_j x+q_j y+\Omega_j t,\quad \Omega_j=p_jq_j,\quad
j=1,2,\cdots,N,\\
&& \exp(A_{ij})=-2(k_i-k_j)^2,\quad i<j=2,3,\cdots, N\\
&& \exp(A_{i,N+j})=-\frac{1}{2(k_i+p_j)^2},\quad
i,j=1,2,\cdots,N,\\
&& \exp(A_{N+i,N+j})=-2(p_i-p_j)^2,\quad i<j=2,3,\cdots,N.
\end{eqnarray}
Here $\alpha_j,\gamma_j$ are both real parameters relating
respectively to the amplitude and phase of the $i$th soliton. The
sum $\sum_{\mu=0,1}^{(e)}$ indicates the summation over all possible
combinations of $\mu_i=0,1$ under the condition
\begin{align}
\sum_{j=1}^N\mu_j=\sum_{j=1}^N\mu_{N+j},\label{mmu1}
\end{align}
and $\sum_{\nu=0,1}^{(o)},\sum_{\lambda=0,1}^{(o)}$ indicate the
summation over all possible combinations of
$\lambda_i=0,1,\nu_i=0,1$ under the condition
\begin{align}
&\sum_{j=1}^N\nu_j=\sum_{j=1}^N\nu_{N+j}+1,\label{mmu2}\\
&\sum_{j=1}^N\lambda_j=\sum_{j=1}^N\lambda_{N+j}-1.\label{mmu3}
\end{align}
The proof of the $N$-soliton solution
here is similar to that of the combined Schr\"odinger-mKdV equation in
\cite{hirota1973} and can be completed by induction. One can check the details therein.

\subsection{Integrable semi-discrete analogue of modified HF equation}
In this section, we construct the integrable discretization of HF
equation \eqref{RS} by using Hirota's discretization method
\cite{Hirota2006}. The Hirota bilinear difference
operator $\exp(\delta D_n)$ is defined
as \cite{AJG},\\
\begin{equation*}
\exp(\delta D_n)a(n)\cdot b(n)\equiv \exp\left
[\delta\Big(\frac{\partial}{\partial n}-\frac{\partial}{\partial
n'}\Big) \right ]a(n)b(n')\mid_{n'=n}=a(n+\delta)b(n-\delta).
\end{equation*}
The differential-difference HF system is obtained by discretizing
the spacial part of the bilinear Eq. \eqref{2HF},
\begin{eqnarray}
&&D_x G\cdot F \rightarrow
\frac{1}{\epsilon}(G_{n+1}F_n-G_nF_{n+1}),\label{d1}\\
&& D_x^2 F\cdot F \rightarrow
\frac{2}{\epsilon^2}\left(F_{n+1}F_{n-1}-F_n^2\right),\label{d2}
\end{eqnarray}
where $x = n\epsilon$, $n$ being integers and $\epsilon$ a
spatial-interval. Substituting \eqref{d1}-\eqref{d2} into
Eq.\eqref{2HF} results to
\begin{eqnarray}
&& D_t G_{n}\cdot F_n +\frac{1}{\epsilon}D_y (G_{n+1}\cdot F_{n}-G_n\cdot F_{n+1})=0, \label{dsc1}\\
&& D_t H_{n}\cdot F_n -\frac{1}{\epsilon}D_y(H_{n+1}\cdot F_{n}-H_n\cdot F_{n+1})=0, \label{dsc2}\\
&&
\frac{2}{\epsilon^2}\left(F_{n+1}F_{n-1}-F_n^2\right)+G_nH_n=0.\label{dsc3}
\end{eqnarray}
We demand that the discretized bilinear forms are invariant under
the gauge transformation:
\begin{eqnarray}
&& F_n \rightarrow F_n \exp(q_0 n),\\
&& G_n \rightarrow G_n \exp(q_0 n),\\
&& H_n \rightarrow H_n \exp(q_0 n).
\end{eqnarray}
Then we find a gauge invariant semi-discrete bilinear HF equations
\begin{eqnarray}
&& (\epsilon D_t+2D_y) G_{n+1}\cdot F_n +(\epsilon D_t-2D_y) G_n\cdot F_{n+1}=0, \label{semb1}\\
&& (\epsilon D_t-2D_y) H_{n+1}\cdot F_n +(\epsilon D_t+2D_y) H_n\cdot F_{n+1}=0, \label{semb2}\\
&&
\frac{2}{\epsilon^2}\left(F_{n+1}F_{n-1}-F_n^2\right)+G_nH_n=0,\label{semb3}
\end{eqnarray}
or equivalently in compact form
\begin{align}
& (\epsilon D_t \cosh\frac{D_n}{2}+2D_y\sinh\frac{D_n}{2})G_{n}\cdot F_n=0, \label{sembb1}\\
& (\epsilon D_t \cosh\frac{D_n}{2}-2D_y\sinh\frac{D_n}{2})H_{n}\cdot F_n=0,  \label{sembb2}\\
& 2(F_{n+1}F_{n-1}- F_n^2)+\epsilon^2G_nH_n=0.\label{sembb3}
\end{align}
Let \begin{align} u_n=\frac{ G_n}{F_n},\quad
v_n=\frac{H_n}{F_n},\quad r_n=\ln \frac{F_{n+1}}{F_n}.\label{dpv}
\end{align} Then Eqs.
\eqref{semb1}-\eqref{semb3} are transformed into ordinary nonlinear
form
\begin{eqnarray}
&& (u_{n+1}+u_{n})_t+(u_{n+1}-u_n)r_{n,t}+\frac{2}{\epsilon}
\Big[(u_{n+1}-u_n)_y+(u_{n+1}+u_n)r_{n,y}\Big]=0,\label{n1}\\
&& (v_{n+1}+v_{n})_t+(v_{n+1}-v_n)r_{n,t}-\frac{2}{\epsilon}
\Big[(v_{n+1}-v_n)_y+(v_{n+1}+v_n)r_{n,y}\Big]=0,\label{n2}\\
&& \frac{2}{\epsilon^2}(e^{r_n-r_{n-1}}-1)+u_{n}v_n=0.\label{n3}
\end{eqnarray}
When we take the continuum limit $\epsilon\rightarrow 0$,
\eqref{n1}-\eqref{n3} reduce to \eqref{non1}-\eqref{non2} and
\begin{align}
2(\ln F)_{xx}+uv=0.\label{n4}
\end{align}
Differentiating Eq.\eqref{n4} with respect to variable $y$, we get
Eq. \eqref{non3}. Thus we regard \eqref{n1}-\eqref{n3} as a
semi-discrete version of the HF system \eqref{HF1}. In the following
discussion we take the interval $\epsilon=1$ for the sake of
simplicity.

Following the Hirota method, we expand $G_n, H_n$ and $F_n$ in series with a small
parameter $\delta$ as
\begin{eqnarray}
&& F_n=1+\delta^2 F_n^{(2)}+\delta^4
F_n^{(4)}+\cdots+\delta^{2k}F_n^{(2k)}+\cdots,
\label{s1}\\
&& G_n=\delta G_n^{(1)}+\delta^3
G_n^{(3)}+\cdots+\delta^{(2k+1)}G_n^{(2k+1)}+\cdots,\label{s2}\\
&& H_n=\delta H_n^{(1)}+\delta^3
H_n^{(3)}+\cdots+\delta^{(2k+1)}H_n^{(2k+1)}+\cdots.\label{s3}
\end{eqnarray}
Substituting the expansion into the above bilinear Eqs.
\eqref{semb1}-\eqref{semb3}, we find that there are only odd order
terms of $\delta$ in the first two equations while only even order
terms appear in the third one. By the standard direct perturbation
method, we obtain the one-soliton solution
\begin{eqnarray}
G_n=\gamma_1\exp(\eta_1), \quad H_n=\gamma_1' \exp(\eta'_1), \quad
F_n=1-\frac{\gamma_1\gamma_1'\beta_1\beta_1'}{(\beta_1\beta'_1-1)^2}\exp(\eta_1+\eta'_1),
\end{eqnarray}
where $\eta_1=p_1 t+q_1y+\ln(\beta_1) n,$ $\eta'_1=p'_1
t+q'_1y+\ln(\beta'_1) n,$ and $\beta_1,\beta_1'$ satisfy
\begin{align}
&\beta_1=\frac{2q_1-p_1}{2q_1+p_1},\\
&\beta_1'=\frac{2q_1'+p_1'}{2q_1'-p_1'},
\end{align}
$\alpha_1, \gamma_1$ and $\alpha'_1, \gamma'_1$ are arbitrary
constants. The two-soliton solution is presented as
\begin{align}
& G_n=\delta [\exp(\eta_1)+\exp(\eta_2)]+\delta^3 [a_{121}
\exp(\eta_1+\eta_2+\eta'_1)
+a_{122}\exp(\eta_1+\eta_2+\eta'_2)],\label{g2st}\\
& H_n=\delta [\exp(\eta_1')+\exp(\eta_2')]+\delta^3 [b_{121}
\exp(\eta_1'+\eta_2'+\eta_1)
+b_{122}\exp(\eta_1'+\eta_2'+\eta_2)],\\
& F_n=1-\delta^2
\Big[\frac{\gamma_1\gamma_1'\beta_1\beta_1'\exp(\eta_1+\eta'_1)}{2(\beta_1\beta'_1-1)^2}
+\frac{\gamma_1\gamma_2'\beta_1\beta_2'\exp(\eta_1+\eta'_2)}{2(\beta_1\beta'_2-1)^2}
+\frac{\gamma_2\gamma_1'\beta_2\beta_1'\exp(\eta_2+\eta_1')}{2(\beta_2\beta'_1-1)^2}\nonumber\\
&\qquad\quad+\frac{\gamma_2\gamma_2'\beta_2\beta_2'\exp(\eta_2+\eta_2')}{2(\beta_2\beta'_2-1)^2}
\Big]+\delta^4 \chi_{1212} [\exp(\eta_1+\eta_2+\eta'_1+\eta'_2)]
\end{align}
where
\begin{align}
&a_{121}=-\frac{\gamma_1\gamma_2\gamma_1'(\beta_1-\beta_2)^2\beta_1'^2}{2(\beta_2\beta_1'-1)^2(\beta_1\beta_1'-1)^2},\quad
a_{122}=-\frac{\gamma_1\gamma_2\gamma_2'(\beta_1-\beta_2)^2\beta_2'^2}{2(\beta_2\beta_2'-1)^2(\beta_1\beta_2'-1)^2},\nonumber\\
&b_{121}=-\frac{\gamma_1\gamma_1'\gamma_2'(\beta'_1-\beta'_2)^2\beta_1^2}{2(\beta_1\beta_1'-1)^2(\beta_1\beta'_2-1)^2},\quad
b_{122}=-\frac{\gamma_2\gamma_1'\gamma_2'(\beta'_1-\beta'_2)^2\beta_2^2}{2(\beta_2\beta_2'-1)^2(\beta_2\beta'_1-1)^2},\nonumber\\
&\chi_{1212}=\frac{(\beta_1-\beta_2)^2(\beta'_1-\beta'_2)^2\gamma_1\gamma_2\gamma_1'\gamma_2'\beta_1\beta_2\beta_1'\beta_2'}
{4(\beta_1\beta'_1-1)^2(\beta_1\beta'_2-1)^2(\beta_2\beta'_1-1)^2(\beta_2\beta'_2-1)^2}.\label{c2t}
\end{align}
We can use the following compact expression for the above
two-soliton solution,
\begin{align}
&F_n=1+a(1,1^*)\gamma_1\gamma_1'\gamma_1'\exp(\eta_1+\eta'_1)+a(1,2^*)\gamma_1\gamma_2'\exp(\eta_1+\eta'_2)\nonumber\\
&\qquad+a(2,1^*)\gamma_2\gamma_1'\exp(\eta_2+\eta'_1)+a(2,2^*)\gamma_2\gamma_2'\exp(\eta_2+\eta'_2)\nonumber\\
&\qquad+a(1,2,1^*,2^*)\gamma_1\gamma_1'\gamma_2\gamma_2'\exp(\eta_1+\eta_2+\eta'_1+\eta'_2),\label{cpt1}\\
&G_n=\gamma_1\exp(\eta_1)+\gamma_2\exp(\eta_2)+a(1,2,1^*)\gamma_1\gamma_2\gamma_1'\exp(\eta_1+\eta_2+\eta'_1)\nonumber\\
&\qquad+a(1,2,2^*)\gamma_1\gamma_2\gamma_2'\exp(\eta_1+\eta_2+\eta'_2),\\
&H_n=\gamma_1'\exp(\eta'_1)+\gamma_2'\exp(\eta'_2)+a(1,1^*,2^*)\gamma_1\gamma_1'\gamma_2'\exp(\eta_1+\eta'_1+\eta'_2)\nonumber\\
&\qquad+a(2,1^*,2^*)\gamma_2\gamma_1'\gamma_2'\exp(\eta_2+\eta'_1+\eta'_2),\label{cpt2}
\end{align}
where the coefficients are defined as
\begin{align}
& a(i,j)=-2\frac{(\beta_i-\beta_j)^2}{\beta_i\beta_j},\label{pfe1}\\
& a(i,j^*)=-\frac{\beta_i\beta_j'}{(\beta_i\beta'_j-1)^2},\label{pfe2}\\
&
a(i^*,j^*)=-2\frac{(\beta'_i-\beta'_j)^2}{\beta'_i\beta'_j},\label{pfe3}
\end{align}
and $a(i,j,k^*),a(i,j^*,k^*),a(i,j,k^*,l^*)$ satisfy the operation
rule \eqref{pfa}. In the same way, we can construct the
three-soliton solution as that in \cite{LG}.  The above expressions
of the one- and two-soliton solutions suggest the exact
$N$-soliton solution of Eqs. \eqref{semb1}-\eqref{semb3} in the
following form
\begin{eqnarray}
&& F_n=\sum_{\mu=0,1}^{(e)}\exp\left[\sum_{j=1}^{N}\mu_j\eta_j
+\sum_{j=N+1}^{2N}\mu_j\eta'_{j-N}+\sum_{1\leq
k<l}^{2N}\mu_k\mu_lA_{kl} \right],\label{solu1}\\
&& G_n=\sum_{\nu=0,1}^{(o)}\exp\left[\sum_{j=1}^{N}\nu_j\eta_j+
\sum_{j=N+1}^{2N}\nu_j\eta'_{j-N}+\sum_{1\leq
k<l}^{2N}\nu_k\nu_lA_{kl}
\right],\label{solu2}\\
&&
H_n=\sum_{\lambda=0,1}^{(o)}\exp\left[\sum_{j=1}^{N}\lambda_j\eta_j+
\sum_{j=N+1}^{2N}\lambda_j\eta'_{j-N}+\sum_{1\leq
k<l}^{2N}\lambda_k\lambda_lA_{kl} \right],\label{solu3}
\end{eqnarray}
where
\begin{eqnarray}
&& \eta_j=p_j t+q_j y+\ln(\beta_j)n+\gamma_j,\quad
\beta_j=\frac{2q_j-p_j}{2q_j+p_j},\quad
j=1,2,\cdots,N,\\
&& \eta'_j=p'_j t+q'_jy+\ln(\beta_j') n+\gamma'_j,\quad
\beta'_j=\frac{2q'_j+p_j'}{2q'_j-p_j'},\quad
j=1,2,\cdots,N,\\
&& \exp(A_{kl})=-2\frac{(\beta_k-\beta_l)^2}{\beta_k\beta_l},\quad k<l=2,3,\cdots, N\\
&&
\exp(A_{k,N+l})=-\frac{1}{2}\frac{\beta_k\beta_l'}{(\beta_k\beta'_l-1)^2},\quad
k,l=1,2,\cdots,N,\\
&&
\exp(A_{N+k,N+l})=-2\frac{(\beta'_k-\beta'_l)^2}{\beta_k'\beta_l'},\quad
k<l=2,3,\cdots,N.
\end{eqnarray}
Here $p_j, q_j, \beta_j, \gamma_j$ are all real parameters. The
summations $\sum_{\mu=0,1}^{(e)},\sum_{\nu=0,1}^{(o)}$ and
$\sum_{\lambda=0,1}^{(o)}$ satisfy the condition
\eqref{mmu1}-\eqref{mmu3} respectively.

\subsection{Soliton propagation and interaction}
By virtue of \eqref{dpv}, \eqref{g2st}-\eqref{c2t}, interaction
between the two solitons can be investigated. From Fig. 1, head-on
elastic interaction between the two solitons is found. With the time
evolution, two solitons  travel towards each other and then
separate. After the interaction, the two solitons maintain their
original amplitudes, velocities and shapes except for the phase
shifts. To interpret the elastic interaction behavior of two
solitons under the condition $\gamma_1\gamma_2\gamma_1'\gamma_2'\neq
0$, asymptotic analysis is carried out in the following.

First we take the notation $\chi_{ij}=a(i,j^*)$ and let
\[
A=\left\{
\left|\eta_{1}+\eta_{1}'\right|,\left|\eta_{1}+\eta_{2}'\right|,
\left|\eta_{2}+\eta_{1}'\right|,\left|\eta_{2}+\eta_{2}'\right|,
\left|\eta_{1}-\eta_{2}\right|,\left|\eta_{1}'-\eta_{2}'\right|\right\}.
\]
When $\gamma_1\gamma_2\gamma_1'\gamma_2'\neq 0$, we have the
following asymptotic behaviors as
\begin{align}
\label{asyp1}
\phi & =\frac{GH}{F^{2}}\nonumber\\
\sim & \begin{cases}
\frac{\gamma_{1}\gamma'_{1}}{\left|\gamma_{1}\gamma'_{1}\chi_{11}\right|}{\rm sech}^{2}\left(\frac{\eta_{1}+\eta_{1}'+\ln\left|\gamma_{1}\gamma'_{1}\chi_{11}\right|}{2}\right) & \eta_{1}+\eta_{1}'\,\,{\rm fixed,}\,\eta_{2}+\eta_{2}'\to-\infty,\left|\eta_{2}+\eta_{2}'\right|=\max A\\
\frac{\gamma_{1}\gamma'_{1}}{\left|\gamma_{1}\gamma'_{1}\chi_{11}\right|}{\rm sech}^{2}\left(\frac{\eta_{1}+\eta_{1}'+\ln\left|\gamma_{1}\gamma'_{1}\chi_{1212}\chi_{22}^{-1}\right|}{2}\right) & \eta_{1}+\eta_{1}'\,\, {\rm fixed,}\,\eta_{2}+\eta_{2}'\to+\infty,\left|\eta_{2}+\eta_{2}'\right|=\max A\\
\frac{\gamma_{2}\gamma'_{2}}{\left|\gamma_{2}\gamma'_{2}\chi_{22}\right|}{\rm sech}^{2}\left(\frac{\eta_{2}+\eta_{2}'+\ln\left|\gamma_{2}\gamma'_{2}\chi_{22}\right|}{2}\right) & \eta_{2}+\eta_{2}'\,\,{\rm fixed,}\,\eta_{1}+\eta_{1}'\to-\infty,\left|\eta_{1}+\eta_{1}'\right|=\max A\\
\frac{\gamma_{2}\gamma'_{2}}{\left|\gamma_{2}\gamma'_{2}\chi_{22}\right|}{\rm sech}^{2}\left(\frac{\eta_{2}+\eta_{2}'+\ln\left|\gamma_{2}\gamma'_{2}\chi_{1212}\chi_{11}^{-1}\right|}{2}\right) & \eta_{2}+\eta_{2}'\,\,{\rm fixed,}\,\eta_{1}+\eta_{1}'\to+\infty,\left|\eta_{1}+\eta_{1}'\right|=\max A\\
\\
\frac{\gamma_{1}\gamma'_{2}}{\left|\gamma_{1}\gamma'_{2}\chi_{12}\right|}{\rm sech}^{2}\left(\frac{\eta_{1}+\eta_{2}'+\ln\left|\gamma_{1}\gamma'_{2}\chi_{12}\right|}{2}\right) & \eta_{1}+\eta_{2}'\,\,{\rm fixed,}\,\eta_{2}+\eta_{1}'\to-\infty,\left|\eta_{2}+\eta_{1}'\right|=\max A\\
\frac{\gamma_{1}\gamma'_{2}}{\left|\gamma_{1}\gamma'_{2}\chi_{12}\right|}{\rm sech}^{2}\left(\frac{\eta_{1}+\eta_{2}'+\ln\left|\gamma_{1}\gamma'_{2}\chi_{1212}\chi_{21}^{-1}\right|}{2}\right) & \eta_{1}+\eta_{2}'\,\,{\rm fixed,}\,\eta_{2}+\eta_{1}'\to+\infty,\left|\eta_{2}+\eta_{1}'\right|=\max A\\
\frac{\gamma_{2}\gamma'_{1}}{\left|\gamma_{2}\gamma'_{1}\chi_{21}\right|}{\rm sech}^{2}\left(\frac{\eta_{2}+\eta_{1}'+\ln\left|\gamma_{2}\gamma'_{1}\chi_{21}\right|}{2}\right) & \eta_{2}+\eta_{1}'\,\,{\rm fixed,}\,\eta_{1}+\eta_{2}'\to-\infty,\left|\eta_{1}+\eta_{1}'\right|=\max A\\
\frac{\gamma_{2}\gamma'_{1}}{\left|\gamma_{2}\gamma'_{1}\chi_{21}\right|}{\rm sech}^{2}\left(\frac{\eta_{2}+\eta_{1}'+\ln\left|\gamma_{2}\gamma'_{1}\chi_{1212}\chi_{12}^{-1}\right|}{2}\right) & \eta_{2}+\eta_{1}'\,\,{\rm fixed,}\,\eta_{1}+\eta_{2}'\to+\infty,\left|\eta_{1}+\eta_{1}'\right|=\max A\\
\\
\frac{\gamma_{1}\gamma{}_{2}a_{122}}{\left|\gamma_{1}\gamma{}_{2}\chi_{12}\chi_{22}\right|}{\rm sech}^{2}\left(\frac{\eta_{1}-\eta_{2}+\ln\left|\gamma_{1}\chi_{12}\gamma_{2}^{-1}\chi_{22}^{-1}\right|}{2}\right) & \eta_{1}-\eta_{2}\,\,{\rm fixed,}\,\eta_{1}'-\eta_{2}'\to-\infty,\left|\eta_{1}'-\eta_{2}'\right|=\max A\\
\frac{\gamma_{1}\gamma{}_{2}a_{121}}{\left|\gamma_{1}\gamma{}_{2}\chi_{11}\chi_{21}\right|}{\rm sech}^{2}\left(\frac{\eta_{1}-\eta_{2}+\ln\left|\gamma_{1}\chi_{11}\gamma_{2}^{-1}\chi_{21}^{-1}\right|}{2}\right) & \eta_{1}-\eta_{2}\,\,{\rm fixed,}\,\eta_{1}'-\eta_{2}'\to+\infty,\left|\eta_{2}+\eta_{1}'\right|=\max A\\
\frac{\gamma'_{1}\gamma'_{2}b_{212}}{\left|\gamma'_{1}\gamma'_{2}\chi_{21}\chi_{22}\right|}{\rm sech}^{2}\left(\frac{\eta_{1}'-\eta_{2}'+\ln\left|\gamma'_{1}\chi_{21}\gamma_{2}'{}^{-1}\chi_{22}^{-1}\right|}{2}\right) & \eta_{1}'-\eta_{2}'\,\,{\rm fixed,}\,\eta_{1}-\eta_{2}\to-\infty,\left|\eta_{1}+\eta_{1}'\right|=\max A\\
\frac{\gamma'_{1}\gamma'_{2}b_{112}}{\left|\gamma'_{1}\gamma'_{2}\chi_{11}\chi_{12}\right|}{\rm
sech}^{2}\left(\frac{\eta_{1}-\eta_{2}+\ln\left|\gamma'_{1}\chi_{11}\gamma'_{2}{}^{-1}\chi_{12}^{-1}\right|}{2}\right)
& \eta_{1}'-\eta_{2}'\,\,{\rm
fixed,}\,\eta_{1}-\eta_{2}\to+\infty,\left|\eta_{1}+\eta_{1}'\right|=\max
A
\end{cases}
\end{align}

\begin{figure}
\begin{center}
\begin{tabular}{cccc}
\includegraphics[scale=0.4]{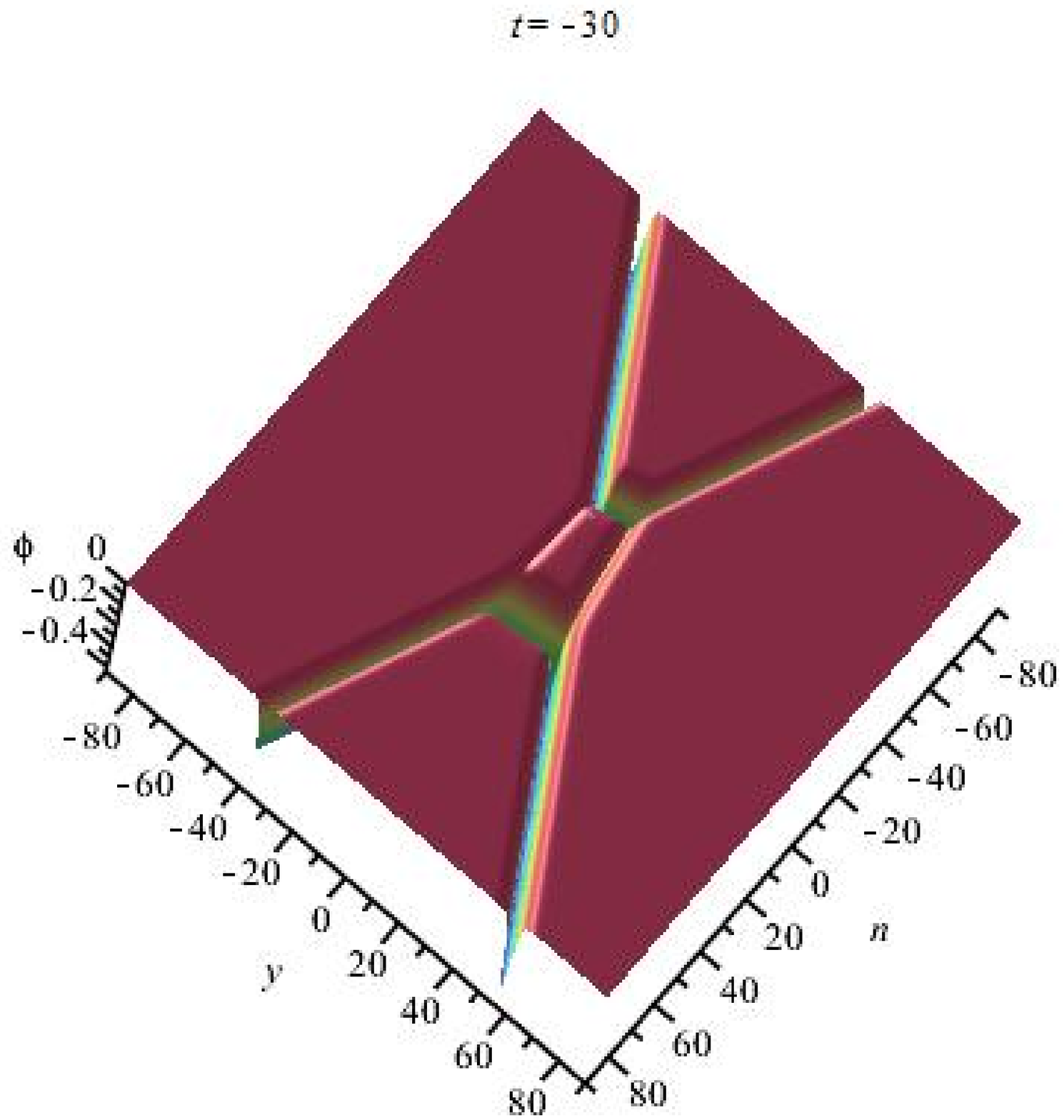}&
\includegraphics[scale=0.4]{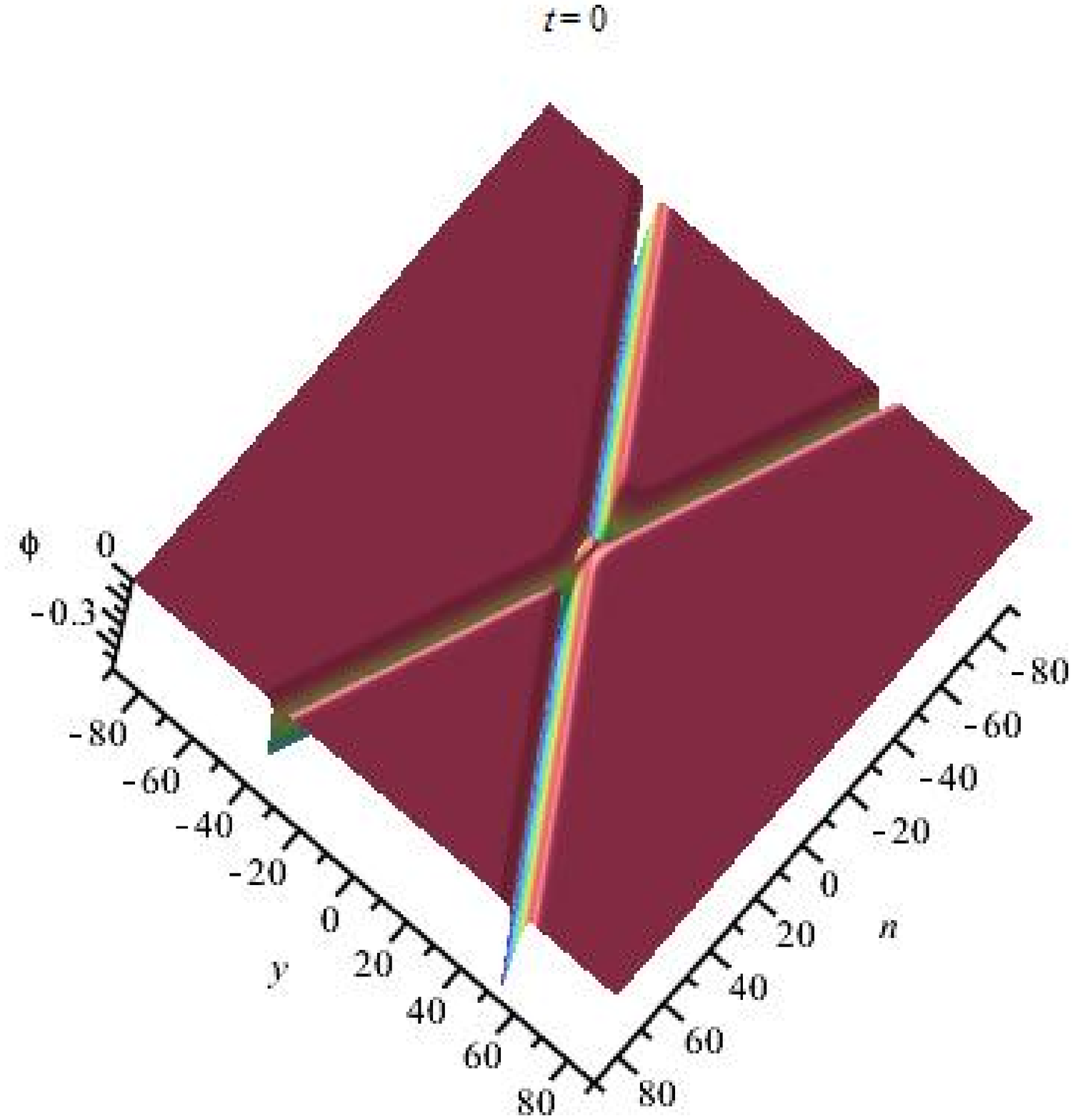}\\
\includegraphics[scale=0.4]{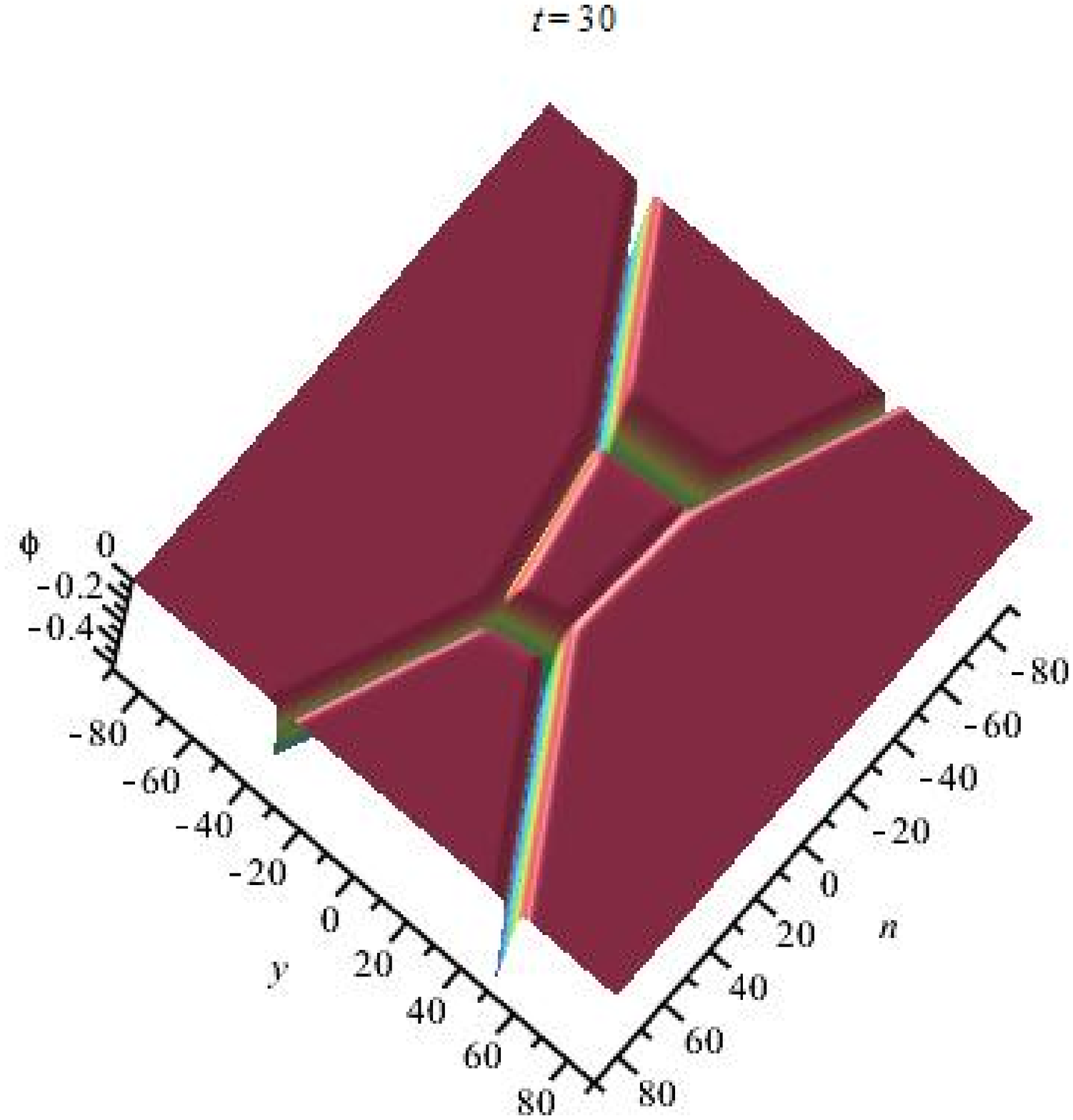} &
\includegraphics[scale=0.4]{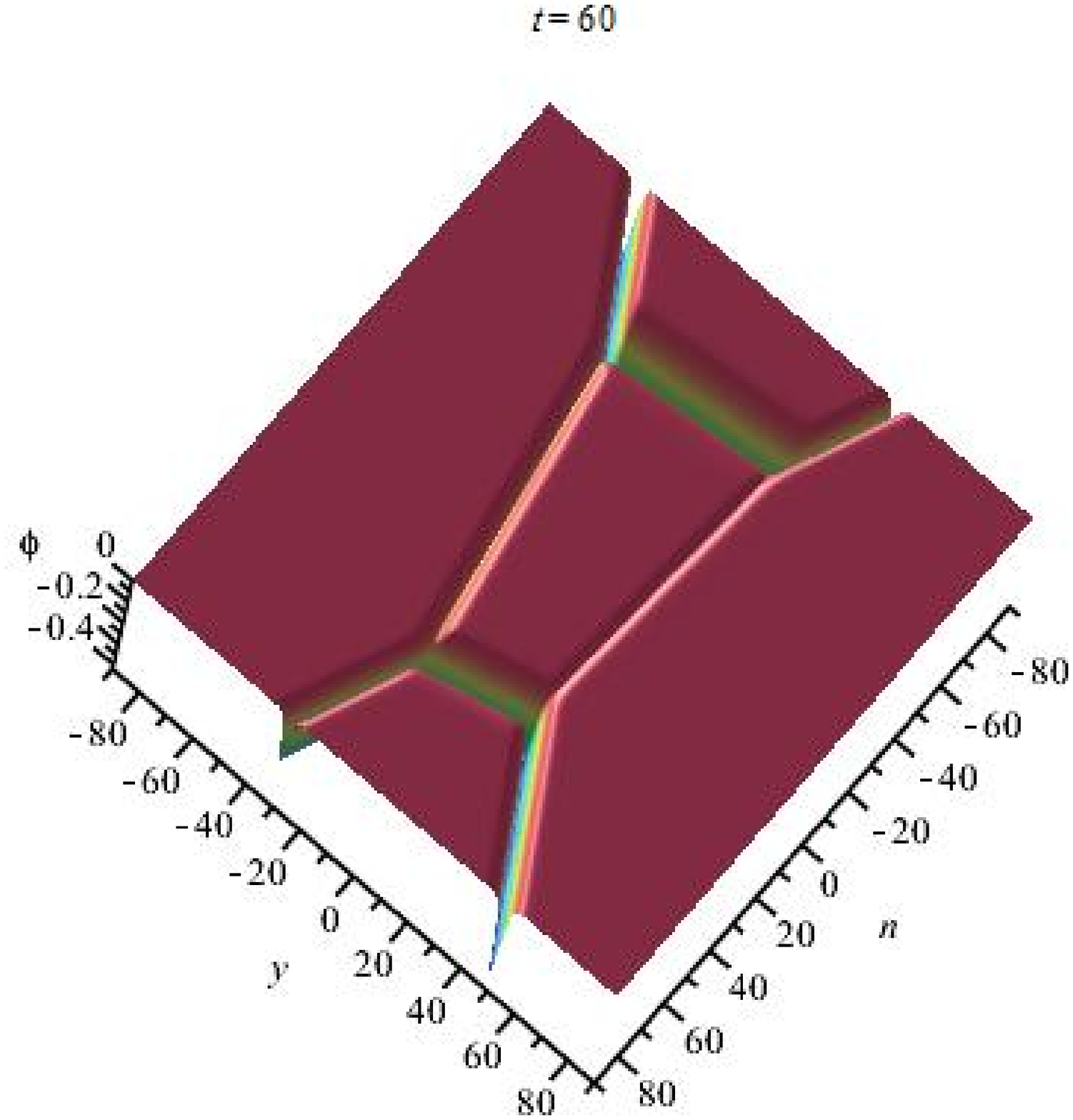}
\end{tabular}
\end{center}
\caption{Two-soliton interaction for $\phi=GH/F^2$ by \eqref{cpt1}-\eqref{cpt2} with parameters
$\gamma_{1}=1,\,\gamma_{2}=1,\,\gamma'_{1}=-1,\,\gamma'_{2}=-1,\,
p_{1}=0.2,\, p_{2}=0.3,q_{1}=-0.7,q_{2}=0.6,\, p'_{1}=0.3,\,
p'_{2}=0.2,\, q'_{1}=-0.8,\, q'_{2}=0.75$}
 \label{Fig1}
\end{figure}

Plots of two solitons interaction  at different times are shown in
Fig. \ref{Fig1}.

In addition, inelastic interaction between the two solitons can be
found. When $\gamma_i$ or $ \gamma_i'$ in \eqref{cpt1}-\eqref{cpt2} is
chosen to be zero, soliton fusion and fission may occur during the
interaction.  Specially, when $\gamma_2'=0$ and
$\gamma_1\gamma_2\gamma_1'\neq0$, we have the following asymptotic
analysis for $\phi$,
\begin{align*}
\phi & = \frac{GH}{F^{2}} =
\frac{\gamma_{1}\gamma'_{1}e^{-\left(\eta_{2}+\eta'_{2}\right)}
+\gamma_{2}\gamma'_{1}e^{-\left(\eta_{1}+\eta'_{2}\right)}+\gamma_{1}\gamma_{2}\gamma_{1}'{}^{2}
a_{121}e^{\eta'_{1}-\eta'_{2}}}{\Big(e^{-\frac{\eta_{1}+\eta_{1}'+\eta_{2}+\eta_{2}'}{2}}
+\chi_{11}\gamma_{1}\gamma'_{1}e^{\frac{\eta_{1}+\eta_{1}'-\eta_{2}-\eta_{2}'}{2}}
+\chi_{21}\gamma_{2}\gamma'_{1}e^{\frac{\eta_{2}+\eta_{1}'-\eta_{1}-\eta_{2}'}{2}}\Big)^{2}}\\
 & \sim \begin{cases}
\frac{\gamma_{1}\gamma'_{1}}{\left|\gamma_{1}\gamma'_{1}\chi_{11}\right|}{\rm
sech}^{2}\left(\frac{\eta_{1}
+\eta_{1}'+\ln\left|\gamma_{1}\gamma'_{1}\chi_{11}\right|}{2}\right)
& \eta_{1}+\eta_{1}'{\, \rm fixed,}\,\eta_{2}
+\eta_{2}'\to-\infty\\
\frac{\gamma_{2}\gamma'_{1}}{\left|\gamma_{2}\gamma'_{1}\chi_{21}\right|}{\rm sech}^{2}\left(\frac{\eta_{2}
+\eta_{1}'+\ln\left|\gamma_{2}\gamma'_{1}\chi_{21}\right|}{2}\right) & \eta_{2}+\eta_{1}'{\,\rm fixed,}\,\eta_{1}+\eta_{2}'\to-\infty\\
\frac{\gamma_{1}\gamma{}_{2}a_{121}}{\left|\gamma_{1}\gamma{}_{2}\chi_{11}\chi_{21}\right|}{\quad\rm
sech}^{2}\Big(\frac{\eta_{1}-\eta_{2}+\ln\left|\gamma_{1}\chi_{11}\gamma_{2}^{-1}\chi_{21}^{-1}\right|}{2}\Big)
& \eta_{1}-\eta_{2}{\,\rm fixed,}\,\eta_{1}'-\eta_{2}'\to+\infty
\end{cases}
\end{align*}
Fig. 2 depicts one soliton split into two solitons with time
evolution under the condition $\gamma_2'=0$.

\begin{figure}
\begin{center}
\begin{tabular}{cccc}
\includegraphics[scale=0.4]{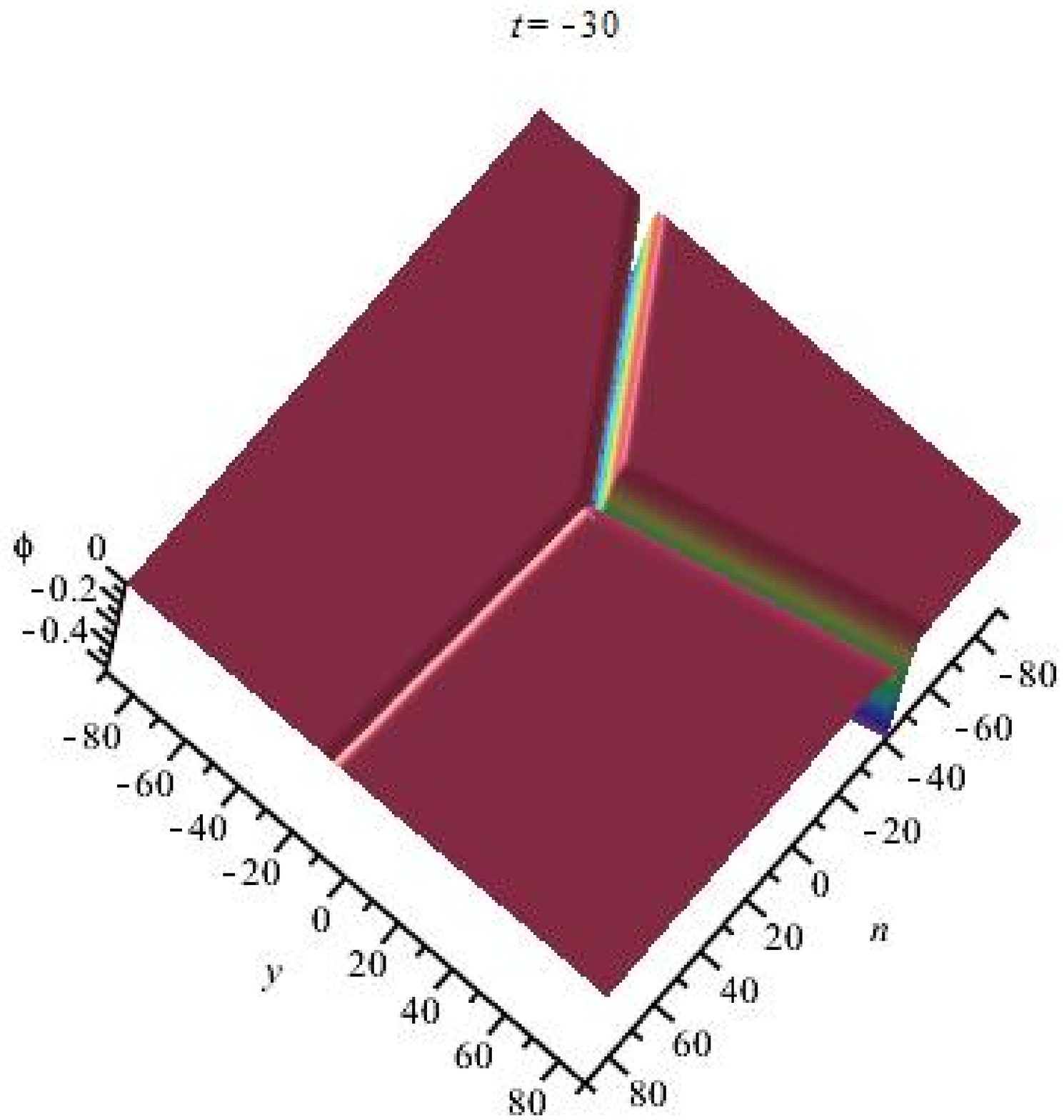}&
\includegraphics[scale=0.4]{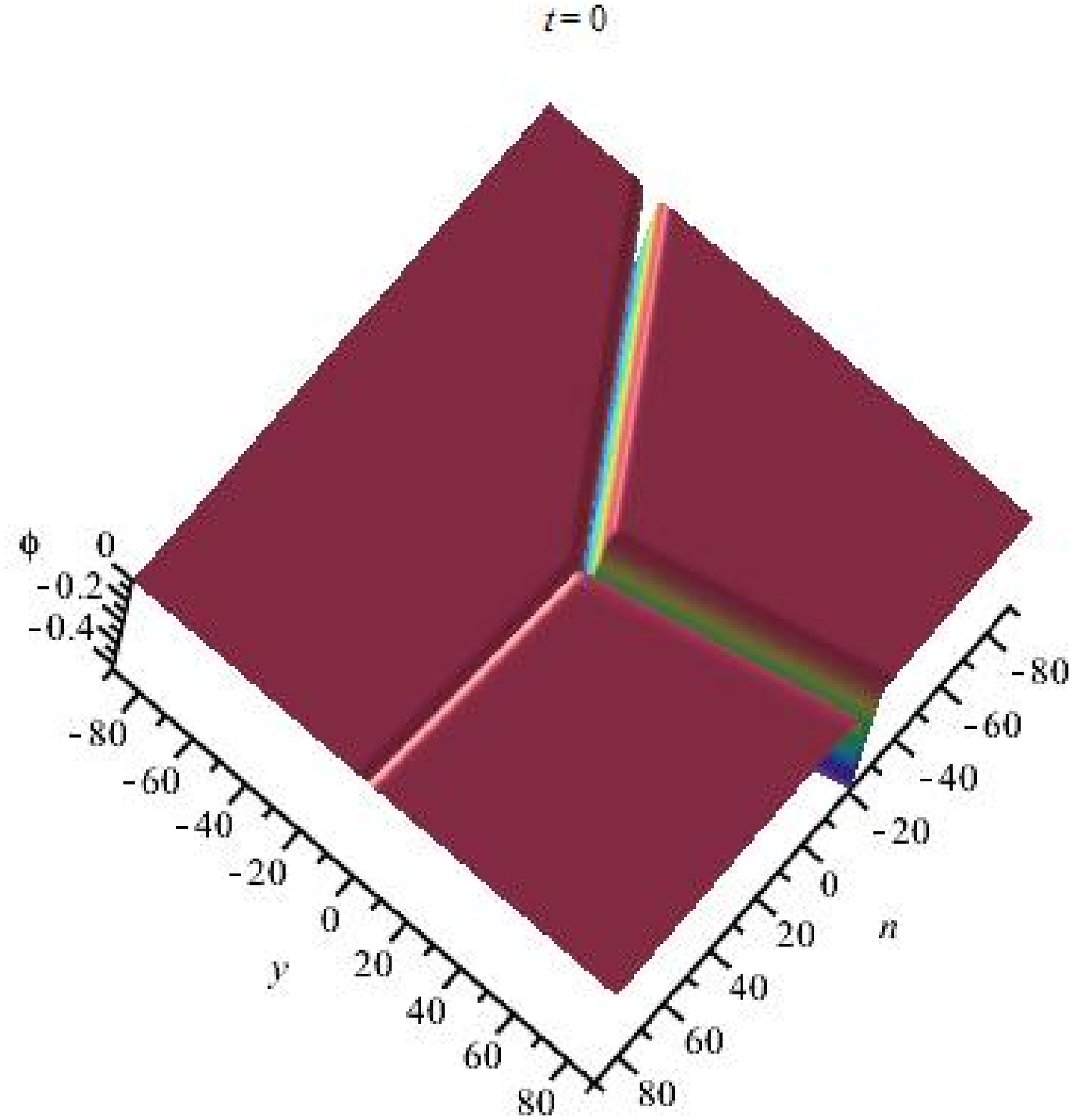}\\
\includegraphics[scale=0.4]{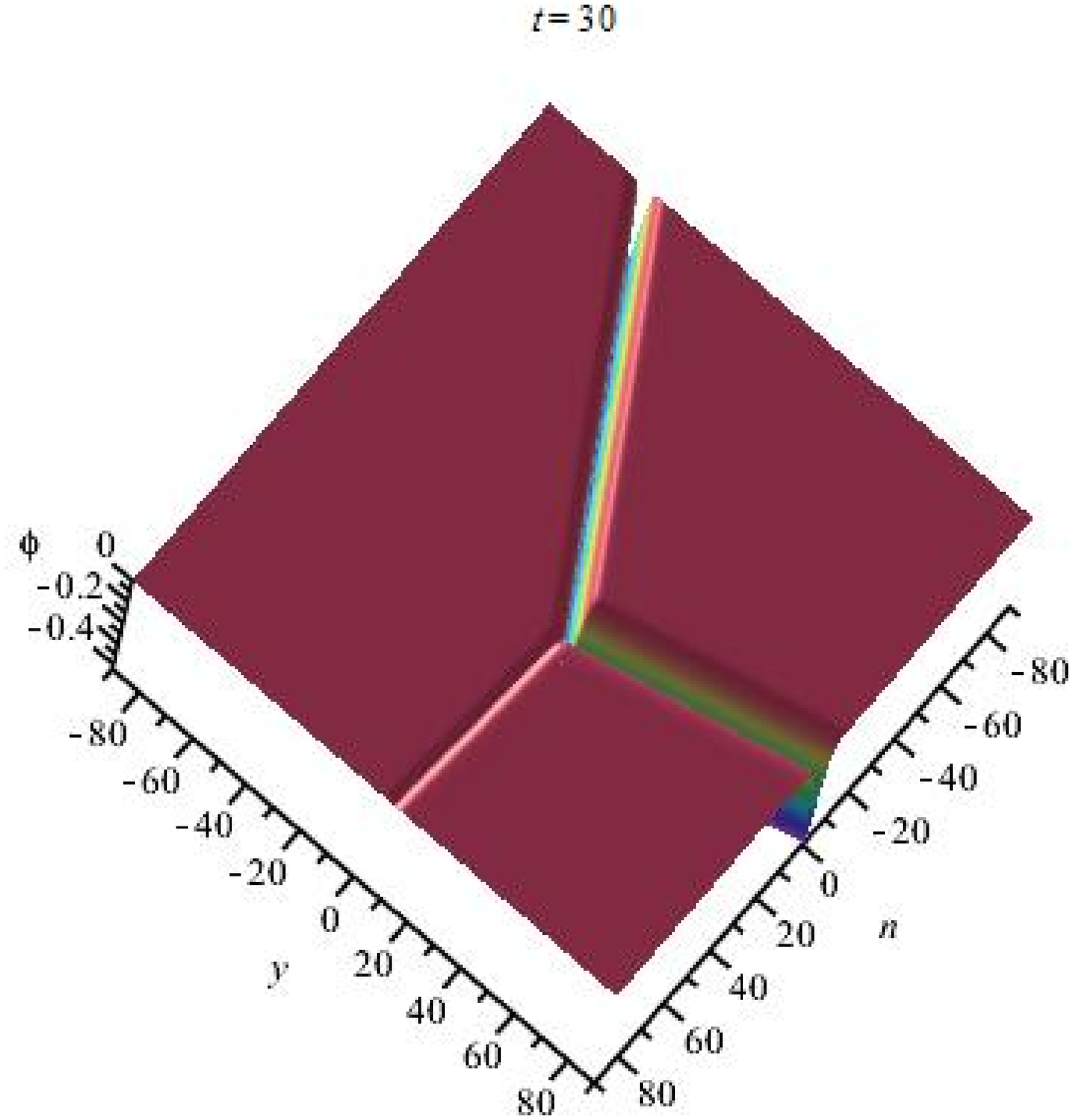}&
\includegraphics[scale=0.4]{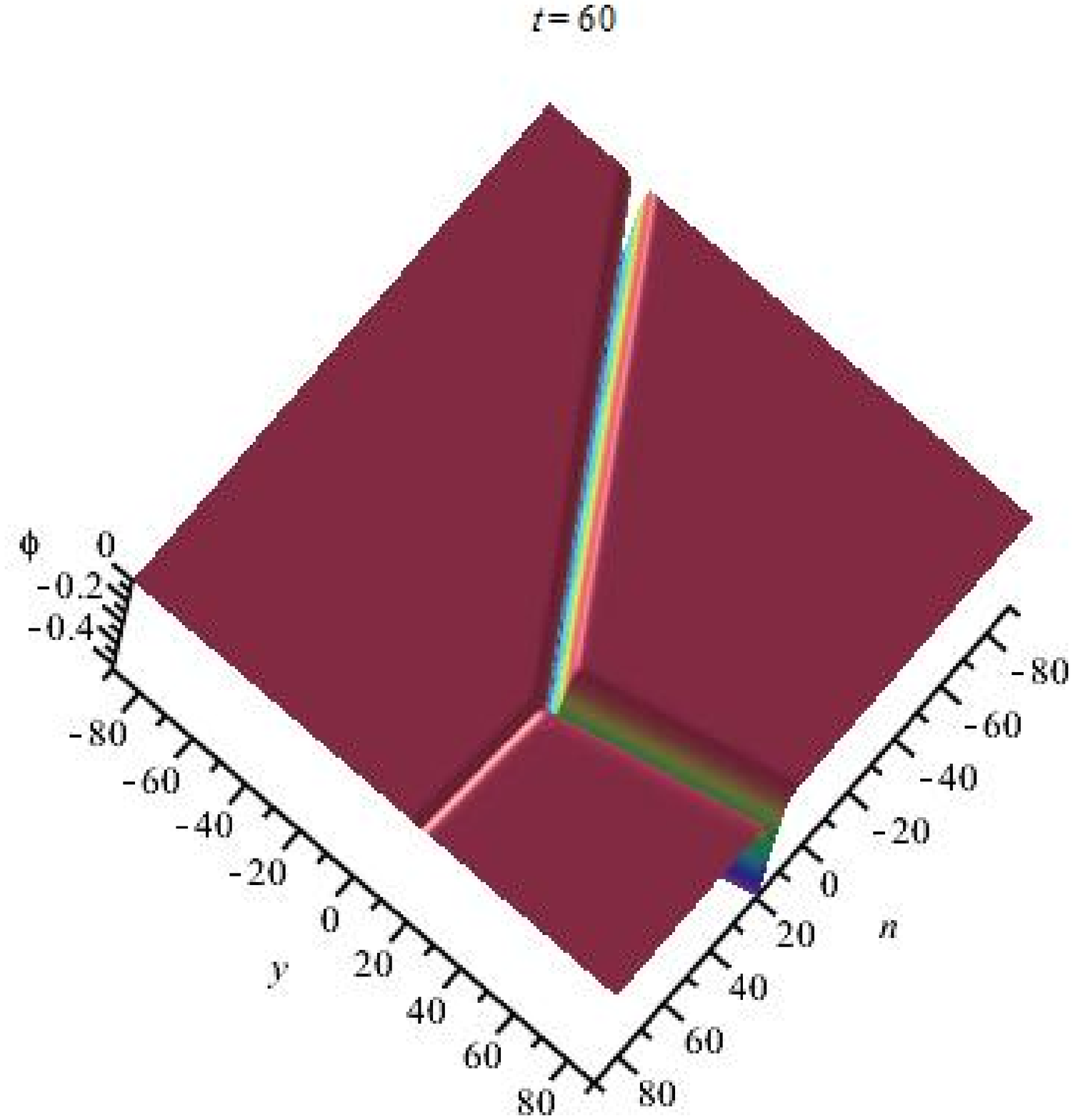}
\end{tabular}
\end{center}
\caption{One soliton splits into two solitons for $\phi=GH/F^2$ by \eqref{cpt1}-\eqref{cpt2} with parameters
$\gamma_{1}=1,\,\gamma_{2}=1,\,\gamma'_{1}=-1,\,\gamma'_{2}=0,\,
p_{1}=0.2,\, p_{2}=0.3,q_{1}=-0.7,q_{2}=0.6,\, p'_{1}=0.3,\,
p'_{2}=0.2,\, q'_{1}=-0.8,\, q'_{2}=0.75$.} \label{Fig2}
\end{figure}

\section{Resonant nonlinear Schr\"odinger equation}

\subsection{multi-soliton solutions of  RNLS equation}
In \cite{TB13}, only two-soliton solution of RNLS equation are
given. In this section, we  present $N-$ soliton solution. We
introduce the transformation
\begin{align}
U=\exp\frac{u + iau + v-iav}{2},
\end{align}
where $u=u(x,t)$ and $v=v(x,t)$ are both real functions, and $a$ is
a real constant.  Eq. \eqref{RS} is transformed into
\begin{align}
&2(i- a)u_t + (1-\beta-a^2 + 2 i a) u_x^2 + 2(1-\beta + i a) u_{xx}
+2(i + a) v_t \nonumber\\
&\qquad+(1 -\beta- a^2-2 i a) v_x^2 + 2(1-\beta- i a) v_{xx} + 2(1
-\beta + a^2) u_x v_x +\alpha \exp(u + v) = 0.\label{mbl}
\end{align}
Since all the functions and parameters in Eq. \eqref{mbl} are real,
separating the real and imaginary parts of Eq. \eqref{mbl} gives as
\begin{align}
& 2 a u_t -(1-\beta-a^2) u_x^2 - 2(1 -\beta) u_{xx}-2 a v_t - (1
-\beta- a^2) v_x^2\nonumber\\
&\qquad- 2(1-\beta) v_{xx}-2(1-\beta + a^2)u_x v_x
-\alpha \exp(u + v) = 0,\label{mb2}\\
& u_t + a u_x^2 + a u_{xx} + v_t - a v_x^2-a v_{xx} = 0.\label{mb3}
\end{align}
In what follows, we set $\beta=1+a^2$. Adding and subtracting
\eqref{mb2} and \eqref{mb3} yield, respectively, two similar
equations,
\begin{align}
& 2(1-a)[u_t+au_{x}^2+au_{xx}]+2(1+a)[v_t-av_x^2-av_{xx} ]+\alpha
e^{u+v}=0,\label{non1}\\
& 2(1+a)[u_t+au_{x}^2+au_{xx}]+2(1-a)[v_t-av_x^2-av_{xx} ]-\alpha
e^{u+v}=0.\label{non2}
\end{align}
It is easy to check that provided Eqs.\eqref{non1}-\eqref{non2} hold
automatically, the following two equations are satisfied,
\begin{align}
& u_{t}+a(u_{xx}+u_x^2)-\frac{\alpha}{4a}e^{u+v}=0,\label{dec1}\\
& v_{t}-a(v_{xx}+v_x^2)+\frac{\alpha}{4a}e^{u+v}=0.\label{dec2}
\end{align}
Through the dependent variable transformation
\begin{align}
u=\ln \frac{g}{f},\qquad v=\ln \frac{h}{f},
\end{align}
Eqs. \eqref{dec1}-\eqref{dec2} are transformed into
\begin{align}
& \frac{4a}{gf}[(D_t+a D_x^2)g\cdot f ]-\frac{1}{f^2}(4a^2
D_x^2f\cdot f+\alpha gh)=0,\label{bn3}\\
& \frac{4a}{hf}[(D_t-a D_x^2)h\cdot f ]+\frac{1}{f^2}(4a^2
D_x^2f\cdot f+\alpha gh)=0,\label{bn4}
\end{align}
that can be decoupled into the following bilinear equations
\begin{subequations}\label{c1}
\begin{eqnarray}
&& (D_t+aD_x^2)g\cdot f=0,\\
&& (D_t-aD_x^2)h\cdot f=0,\\
&& 4a^2D_x^2f\cdot f+\alpha gh=0.
\end{eqnarray}
\end{subequations}

We remark here that the bilinear equations \eqref{c1} are presented
in \cite{TB13} by using binary Bell polynomials method. To find
multi-soliton solutions of Eqs. \eqref{c1}, we expand $f,g$ and $h$
as following
\begin{align}
& f=1+\epsilon^2 f^{(2)}+\epsilon^4 f^{(4)}+\cdots,\label{exp1}\\
& g=\epsilon g^{(1)}+\epsilon^3 g^{(3)}+\cdots,\\
& h=\epsilon h^{(1)}+\epsilon h^{(3)}+\cdots.\label{exp2}
\end{align}
Substituting the expansion \eqref{exp1}-\eqref{exp2} into the
bilinear Eq. \eqref{c1} we find that there are only odd order terms
of $\epsilon$ in the first two equations while only even order terms
appear in the third one. Comparing the coefficients at each order of
$\epsilon$, we obtain one-soliton solution
\begin{align}
& g=\epsilon \exp(\eta),\quad h=\epsilon \exp(\xi),\\
& f=1+\epsilon^2 \frac{-\alpha}{8a^2(p+r)^2}\exp(\eta+\xi),
\end{align}
where $\eta=px+qt+\eta^0,\xi=rx+st+\xi^0$ and $q,s$ satisfy the
dispersion relation
\begin{align}
q=-ap^2,\qquad s=ar^2.
\end{align}
Here $p,r$ and $\eta^0,\xi^0$ are arbitrary constants. The
two-soliton solution is presented as follows
\begin{align}
& g=\epsilon
\Big(\exp(\eta_1)+\exp(\eta_2)\Big)+\epsilon^3\Big(a_{121}\exp(\eta_1+\eta_2+\xi_1)
+a_{122}\exp(\eta_1+\eta_2+\xi_2)\Big),\\
& h=\epsilon\Big
(\exp(\xi_1)+\exp(\xi_2)\Big)+\epsilon^3\Big(b_{121}\exp(\xi_1+\xi_2+\eta_1)
+b_{122}\exp(\xi_1+\xi_2+\eta_2)\Big),\\
&
f=1+\epsilon^2\Big[c_{11}\exp(\eta_1+\xi_1)+c_{12}\exp(\eta_1+\xi_2)
+c_{21}\exp(\eta_2+\xi_1)+c_{22}\exp(\eta_2+\xi_2)\Big]\nonumber\\
&\qquad+\epsilon^4c_{1212}\exp(\eta_1+\eta_2+\xi_1+\xi_2),
\end{align}
where
\begin{align}
& \eta_i=p_ix+q_it+\eta_i^0,\quad \xi_j=r_jx+s_jt+\xi_j^0,\\
& q_i=-ap_i^2,\quad s_j=ar_j^2,\\
& a_{121}=-\frac{\alpha
(p_1-p_2)^2}{8a^2(p_1+r_1)^2(p_2+r_1)^2},\quad a_{122}=-\frac{\alpha
(p_1-p_2)^2}{8a^2(p_1+r_2)^2(p_2+r_2)^2},\\
& b_{121}=-\frac{\alpha
(r_1-r_2)^2}{8a^2(p_1+r_1)^2(p_1+r_2)^2},\quad b_{122}=-\frac{\alpha
(r_1-r_2)^2}{8a^2(p_2+r_1)^2(p_2+r_2)^2},\\
& c_{ij}=-\frac{\alpha}{8a^2(p_i+k_j)^2},\quad i,j=1,2,\\
& c_{1212}=\frac{\alpha^2
(p_1-p_2)^2(r_1-r_2)^2}{64a^4(p_1+r_1)(p_1+r_2)(p_2+r_1)(p_2+r_2)}.
\end{align}
We can use the following compact expression for the two-soliton
solution,
\begin{align}
&g=\exp(\eta_1)+\exp(\eta_2)+a(1,2,1^*)\exp(\eta_1+\eta_2+\xi_1)\nonumber\\
&\qquad+a(1,2,2^*)\exp(\eta_1+\eta_2+\eta_2^*),\label{2st1}\\
&h=\exp(\xi_1)+\exp(\xi_2)+a(1,1^*,2^*)\exp(\eta_1+\xi_1+\xi_2)\nonumber\\
&\qquad+a(2,1^*,2^*)\exp(\eta_2+\xi_1+\xi_2),\\
&f=1+a(1,1^*)\exp(\eta_1+\xi_1)+a(1,2^*)\exp(\eta_1+\xi_2)\nonumber\\
&\qquad+a(2,1^*)\exp(\eta_2+\xi_1)+a(2,2^*)\exp(\eta_2+\xi_2)\nonumber\\
&\qquad+a(1,2,1^*,2^*)\exp(\eta_1+\eta_2+\xi_1+\xi_2),\label{2st2}
\end{align}
where the coefficients are defined as
\begin{align}
& a(i,j^*)=-\frac{\alpha}{8a^2(p_i+r_j)},\label{ree1}\\
& a(i,j)=-\frac{8a^2(p_i-p_j)^2}{\alpha},\\
& a(i^*,j^*)=-\frac{8a^2(r_i-r_j)^2}{\alpha},\\
& a(i_1,i_2,\cdots,i_n)=\prod_{1\leq k<l\leq n} a(i_k,i_l).\label{ree2}
\end{align}

In the same way, we give the exact $N$-soliton solution of Eqs. \eqref{c1} in the
following form
\begin{eqnarray}
&& f=\sum_{\mu=0,1}^{(e)}\exp\left[\sum_{j=1}^{N}\mu_j\eta_j
+\sum_{j=N+1}^{2N}\mu_j\xi_{j-N}+\sum_{1\leq
k<l}^{2N}\mu_k\mu_lA_{kl} \right],\label{solu11}\\
&& g=\sum_{\nu=0,1}^{(o)}\exp\left[\sum_{j=1}^{N}\nu_j\eta_j+
\sum_{j=N+1}^{2N}\nu_j\xi_{j-N}+\sum_{1\leq
k<l}^{2N}\nu_k\nu_lA_{kl}
\right],\label{solu12}\\
&&
h=\sum_{\lambda=0,1}^{(o)}\exp\left[\sum_{j=1}^{N}\lambda_j\eta_j+
\sum_{j=N+1}^{2N}\lambda_j\xi_{j-N}+\sum_{1\leq
k<l}^{2N}\lambda_k\lambda_lA_{kl} \right],\label{solu13}
\end{eqnarray}
where
\begin{eqnarray}
&& \eta_i=p_ix+q_it+\eta_j^0,\quad q_j=-ap_j^2, \quad
j=1,2,\cdots,N,\\
&& \xi_j=r_jx+s_jt+\xi_j^0,\quad s_j=ar_j^2,\quad
j=1,2,\cdots,N,\\
&& \exp(A_{kl})=-\frac{8a^2(p_k-p_l)^2}{\alpha},\quad k<l=2,3,\cdots, N\\
&& \exp(A_{k,N+l})=-\frac{\alpha}{8a^2(p_k+r_l)},\quad
k,l=1,2,\cdots,N,\\
&& \exp(A_{N+k,N+l})=-\frac{8a^2(r_k-r_l)^2}{\alpha},\quad
k<l=2,3,\cdots,N.\end{eqnarray}
Here $p_j,r_j$ are both real parameters relating respectively to the
amplitude and phase of the $i$th soliton. The
summations $\sum_{\mu=0,1}^{(e)},\sum_{\nu=0,1}^{(o)}$ and
$\sum_{\lambda=0,1}^{(o)}$ satisfy the condition
\eqref{mmu1}-\eqref{mmu3} respectively.

\subsection{Integrable discretization of  RNLS equation}
The differential-difference RNLS equation is
obtained by discretizing the spacial part of the bilinear Eq.
\eqref{c1},
\begin{eqnarray}
D_x^2 g\cdot f \rightarrow
\frac{1}{\delta^2}(g_{n+1}f_{n-1}-2g_nf_n+g_{n-1}f_{n+1}),
\end{eqnarray}
where $x = n\delta$, $n$ being integers and $\delta$ a
spatial-interval. We get
\begin{eqnarray}
&& D_t g_{n}\cdot f_n +\frac{a}{\delta^2}(g_{n+1}f_{n-1}-2g_nf_n+g_{n-1}f_{n+1})=0, \label{b3}\\
&& D_t h_{n}\cdot f_n -\frac{a}{\delta^2}(h_{n+1}f_{n-1}-2h_nf_n+h_{n-1}f_{n+1})=0, \label{b4}\\
&& \frac{8a^2}{\delta^2}(f_{n+1}f_{n-1}-f_n^2)+\alpha
g_nh_n=0.\label{b5}
\end{eqnarray}
Let $g_n=e^{u_n} f_n,\, h_n=e^{v_n}f_n$. Then Eqs.
\eqref{b3}-\eqref{b5} transforms into the following nonlinear form
\begin{eqnarray}
&& u_{n,t}+\frac{a}{\delta^2}\Big[ (1-\frac{\alpha\delta^2}{8a^2}
e^{u_n+v_n})(e^{u_{n+1}-u_{n}}+e^{u_{n-1}-u_n})-2
\Big]=0,\label{n1}\\
&& v_{n,t}-\frac{a}{\delta^2}\Big[ (1-\frac{\alpha\delta^2}{8a^2}
e^{u_n+v_n})(e^{v_{n+1}-v_{n}}+e^{v_{n-1}-v_n})+2 \Big]=0.\label{n2}
\end{eqnarray}
When we take the continuum limit $\delta\rightarrow 0$, Eqs.
\eqref{n1}-\eqref{n2} tend to Eqs. \eqref{dec1}-\eqref{dec2}. In the
sequel we use this discrete spacial step $\delta=1$. Multiplying
\eqref{n1} by $2(1-a)(e^{v_{n+1}-v_n}+e^{v_{n-1}-v_n})$, and
\eqref{n2} by $2(1+a)(e^{u_{n+1}-u_n}+e^{u_{n-1}-u_n})$, adding and
subtracting each other, respectively, yield
\begin{align}
& 2(1+a)\Big[u_{n,t}+a(e^{u_{n+1}-u_n}+e^{u_{n-1}-u_n}-2)\Big]
(e^{v_{n+1}-v_n}+e^{v_{n-1}-v_n})\nonumber\\
&\quad+2(1-a)\Big[v_{n,t}-a(e^{v_{n+1}-v_n}+e^{v_{n-1}-v_n}+2)\Big]
(e^{u_{n+1}-u_n}+e^{u_{n-1}-u_n})\nonumber\\
&\quad-\frac{\alpha}{2}e^{u_n+v_n}(e^{u_{n+1}-u_n}+e^{u_{n-1}-u_n})
(e^{v_{n+1}-v_n}+e^{v_{n-1}-v_n})=0,\label{dn1}\\
& 2(1-a)\Big[u_{n,t}+a(e^{u_{n+1}-u_n}+e^{u_{n-1}-u_n}-2)\Big]
(e^{v_{n+1}-v_n}+e^{v_{n-1}-v_n})\nonumber\\
&\quad+2(1+a)\Big[v_{n,t}-a(e^{v_{n+1}-v_n}+e^{v_{n-1}-v_n}+2)\Big]
(e^{u_{n+1}-u_n}+e^{u_{n-1}-u_n})\nonumber\\
&\quad+\frac{\alpha}{2}e^{u_n+v_n}(e^{u_{n+1}-u_n}+e^{u_{n-1}-u_n})
(e^{v_{n+1}-v_n}+e^{v_{n-1}-v_n})=0. \label{dn2}
\end{align}
Adding  and subtracting \eqref{dn1}-\eqref{dn2} give us
\begin{align}
&\Big[u_{n,t}+a(e^{u_{n+1}-u_n}+e^{u_{n-1}-u_n}-2)\Big]
(e^{v_{n+1}-v_n}+e^{v_{n-1}-v_n})\nonumber\\
&\quad+\Big[v_{n,t}-a(e^{v_{n+1}-v_n}+e^{v_{n-1}-v_n}+2)\Big]
(e^{u_{n+1}-u_n}+e^{u_{n-1}-u_n})=0,\label{dn3}\\
&4a\Big[u_{n,t}+a(e^{u_{n+1}-u_n}+e^{u_{n-1}-u_n}-2)\Big]
(e^{v_{n+1}-v_n}+e^{v_{n-1}-v_n})\nonumber\\
&\quad-4a\Big[v_{n,t}-a(e^{v_{n+1}-v_n}+e^{v_{n-1}-v_n}+2)\Big]
(e^{u_{n+1}-u_n}+e^{u_{n-1}-u_n})\nonumber\\
&\quad-\alpha e^{u_n+v_n}(e^{u_{n+1}-u_n}+e^{u_{n-1}-u_n})
(e^{v_{n+1}-v_n}+e^{v_{n-1}-v_n})=0.\label{dn4}
\end{align}
From \eqref{dn3} and \eqref{dn4}, we get
\begin{align}
&4(i-a)\Big[u_{n,t}+a(e^{u_{n+1}-u_n}+e^{u_{n-1}-u_n}-2)\Big]
(e^{v_{n+1}-v_n}+e^{v_{n-1}-v_n})\nonumber\\
&\quad+4(a+i)\Big[v_{n,t}-a(e^{v_{n+1}-v_n}+e^{v_{n-1}-v_n}+2)\Big]
(e^{u_{n+1}-u_n}+e^{u_{n-1}-u_n})\nonumber\\
&\quad+\alpha e^{u_n+v_n}(e^{u_{n+1}-u_n}+e^{u_{n-1}-u_n})
(e^{v_{n+1}-v_n}+e^{v_{n-1}-v_n})=0.\label{sdrnls}
\end{align}
Setting
\begin{align}
U_n=\exp{\Big(\frac{u_n+v_n}{2}+ia\frac{u_n-v_n}{2}\Big)},\quad
V_n=U_n^*=\exp{\Big(\frac{u_n+v_n}{2}-ia\frac{u_n-v_n}{2}\Big)},
\end{align}
or equivalently,
\begin{align}
u_n=\ln(U_n^{\frac{a-i}{2a}}V_n^{\frac{a+i}{2a}}),\quad
v_n=\ln(U_n^{\frac{a+i}{2a}}V_n^{\frac{a-i}{2a}}),
\end{align}
and substituting $u_n,v_n$ by $U_n, V_n$ into Eq. \eqref{sdrnls}, we
obtain a semi-discrete RNLS equation
\begin{align}
&4(i-a)\Big[\mathrm{Re}\Big(\frac{a-i}{a}\frac{U_{n,t}}{U_n}\Big)
+a\Big(\Big|\Big(\frac{U_{n+1}}{U_n}\Big)^{\frac{a-i}{2a}}\Big|^2
+\Big|\Big(\frac{U_{n-1}}{U_n}\Big)^{\frac{a-i}{2a}}\Big|^2-2\Big)\Big]\Big[
\Big|\Big(\frac{U_{n+1}}{U_n}\Big)^{\frac{a+i}{2a}}\Big|^2
+\Big|\Big(\frac{U_{n-1}}{U_n}\Big)^{\frac{a+i}{2a}}\Big|^2\Big]\nonumber\\
&+4(i+a)\Big[\mathrm{Re}\Big(\frac{a+i}{2a}\frac{U_{n,t}}{U_n}\Big)
-a\Big(\Big|\Big(\frac{U_{n+1}}{U_n}\Big)^{\frac{a+i}{2a}}\Big|^2
+\Big|\Big(\frac{U_{n-1}}{U_n}\Big)^{\frac{a+i}{2a}}\Big|^2+2\Big)\Big]\Big[
\Big|\Big(\frac{U_{n+1}}{U_n}\Big)^{\frac{a-i}{2a}}\Big|^2
+\Big|\Big(\frac{U_{n-1}}{U_n}\Big)^{\frac{a-i}{2a}}\Big|^2\Big]\nonumber\\
&+\alpha \Big|U_n\Big|^2
\Big(\Big|\Big(\frac{U_{n+1}}{U_n}\Big)^{\frac{a+i}{2a}}\Big|^2
+\Big|\Big(\frac{U_{n-1}}{U_n}\Big)^{\frac{a+i}{2a}}\Big|^2\Big)
\Big(\Big|\Big(\frac{U_{n+1}}{U_n}\Big)^{\frac{a-i}{2a}}\Big|^2
+\Big|\Big(\frac{U_{n-1}}{U_n}\Big)^{\frac{a-i}{2a}}\Big|^2\Big)=0.\label{semirnls}
\end{align}
In what follows we put $a=1$ and $\alpha=-8$ for the sake of
simplicity. With these special parameters, the semi-discrete
equations \eqref{n1}-\eqref{n2} are simplified as follows
\begin{eqnarray}
&& u_{n,t}+(1+ e^{u_n+v_n})(e^{u_{n+1}-u_{n}}+e^{u_{n-1}-u_n})-2
=0,\label{n3}\\
&& v_{n,t}-(1+ e^{u_n+v_n})(e^{v_{n+1}-v_{n}}+e^{v_{n-1}-v_n})+2
=0.\label{n4}
\end{eqnarray}
By the transformation $e^{u_n}\rightarrow u_n$ and
$e^{v_n}\rightarrow v_n$, Eqs. \eqref{n3}-\eqref{n4} can be
expressed in an alternative simpler form
\begin{eqnarray}
&& u_{n,t}-2u_n+(1+ u_nv_n)(u_{n+1}+u_{n-1})
=0,\label{n5}\\
&& v_{n,t}+2v_n-(1+ u_nv_n)(v_{n+1}+v_{n-1})=0.\label{n6}
\end{eqnarray}

\subsection{Soliton solutions of the integrable semi-discrete version of RNLS}
In this section, we construct the multi-soliton solutions of the
semi-discrete RNLS equation \eqref{semirnls} and hence show its
integrability. We rewrite Eqs. \eqref{b3}-\eqref{b5} with
$\delta=a=1$ in following compact form
\begin{align}
& \Big(D_t-2+2\cosh(D_n\Big)g_n\cdot f_n=0, \label{sembb1}\\
& \Big(D_t+2-2 \cosh(D_n)\Big)h_{n}\cdot f_n=0,  \label{sembb2}\\
& 2\sinh^2\left(\frac{D_n}{2}\right) f_n\cdot
f_n-g_nh_n=0.\label{sembb3}
\end{align}

Similar to the continuous case, we expand $f_n,g_n$
and $h_n$ into series with respect to a small parameter $\epsilon$
as follows
\begin{eqnarray}
&& f_n=1+\epsilon^2 f_n^{(2)}+\epsilon^4
f_n^{(4)}+\cdots+\epsilon^{2k}f_n^{(2k)}+\cdots,
\label{s1}\\
&& g_n=\epsilon g_n^{(1)}+\epsilon^3
g_n^{(3)}+\cdots+\epsilon^{2k+1}g_n^{(2k+1)}+\cdots,\label{s2}\\
&& h_n=\epsilon h_n^{(1)}+\epsilon^3
h_n^{(3)}+\cdots+\epsilon^{2k+1}h_n^{(2k+1)}+\cdots.\label{s3}
\end{eqnarray}
We obtain the one-soliton solution
\begin{eqnarray}
g_n=\epsilon\exp(\eta_1), \quad h_n=\epsilon \exp(\eta'_1), \quad
f_n=1+
\frac{\epsilon^2}{4}\csch^2(\frac{\beta_1+\beta'_1}{2})\exp(\eta_1+\eta'_1),
\end{eqnarray}
where $\eta_1=\alpha_1 t+\beta_1 n+\gamma_1,$ $\eta'_1=\alpha'_1
t+\beta'_1 n+\gamma'_1,$ and $\beta_1,\beta_1'$ satisfy the
dispersion relation
\begin{align}
&\alpha_1+4\sinh^2(\frac{\beta_1}{2})=0,\\
&\alpha_1'-4\sinh^2(\frac{\beta_1'}{2})=0.
\end{align}
Here $\alpha_1, \gamma_1$ and $\alpha'_1, \gamma'_1$ are arbitrary
parameters. The two-soliton solution is presented as follows
\begin{align}
& g_n=\epsilon [\exp(\eta_1)+\exp(\eta_2)]+\epsilon^3 [\varrho_1
\exp(\eta_1+\eta_2+\eta'_1)
+\varrho_2\exp(\eta_1+\eta_2+\eta'_2)],\\
& h_n=\epsilon [\exp(\eta_1')+\exp(\eta_2')]+\epsilon^3 [\varsigma_1
\exp(\eta_1'+\eta_2'+\eta_1)
+\varsigma_2\exp(\eta_1'+\eta_2'+\eta_2)],\\
& f_n=1+\epsilon^2
\Big[\frac{1}{4}\csch^2(\frac{\beta_1+\beta'_1}{2})\exp(\eta_1+\eta'_1)
+\frac{1}{4}\csch^2(\frac{\beta_1+\beta'_2}{2})\exp(\eta_1+\eta'_2)\nonumber\\
&\qquad+\frac{1}{4}\csch^2(\frac{\beta_2+\beta'_1}{2})\exp(\eta_2+\eta_1')
+\frac{1}{4}\csch^2(\frac{\beta_2+\beta'_2}{2})\exp(\eta_2+\eta_2')
\Big]\nonumber\\
&\quad\quad+\epsilon^4 \chi [\exp(\eta_1+\eta_2+\eta'_1+\eta'_2)],
\end{align}
where
\begin{align}
&\varrho_1=\frac{(e^{\beta_1+\beta'_1}-e^{\beta_2+\beta'_1})^2}
{(e^{\beta_1+\beta'_1}-1)^2(e^{\beta_2+\beta'_1}-1)^2},\quad
\varrho_2=\frac{(e^{\beta_1+\beta'_2}-e^{\beta_2+\beta'_2})^2}
{(e^{\beta_1+\beta'_2}-1)^2(e^{\beta_2+\beta'_2}-1)^2},\nonumber\\
&\varsigma_1=\frac{(e^{\beta_1+\beta'_1}-e^{\beta_1+\beta'_2})^2}
{(e^{\beta_1+\beta'_1}-1)^2(e^{\beta_1+\beta'_2}-1)^2},\quad
\varsigma_2=\frac{(e^{\beta_2+\beta'_1}-e^{\beta_2+\beta'_2})^2}
{(e^{\beta_2+\beta'_1}-1)^2(e^{\beta_2+\beta'_2}-1)^2},\nonumber\\
&\chi=\frac{(e^{\beta'_1}-e^{\beta'_2})^2(e^{\beta_1}-e^{\beta_2})^2
e^{\beta_1+\beta_2 +\beta'_1+\beta'_2}}
{(e^{\beta_2+\beta'_1}-1)^2(e^{\beta_2+\beta'_2}-1)^2
(e^{\beta_1+\beta'_1}-1)^2(e^{\beta_1+\beta'_2}-1)^2}.
\end{align}

\begin{figure}
\begin{center}
\begin{tabular}{cccc}
\includegraphics[scale=0.4]{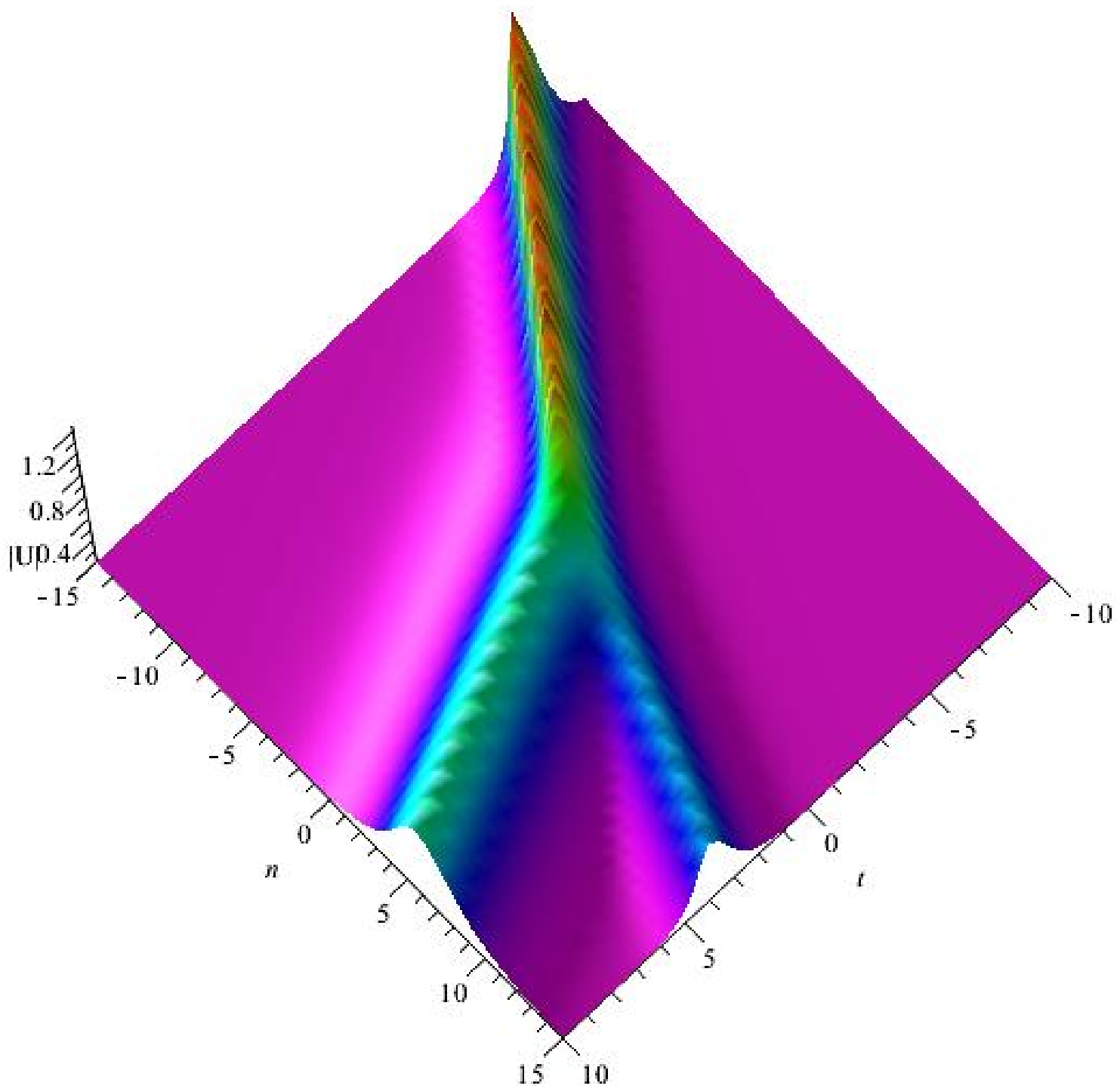}&
\includegraphics[scale=0.4]{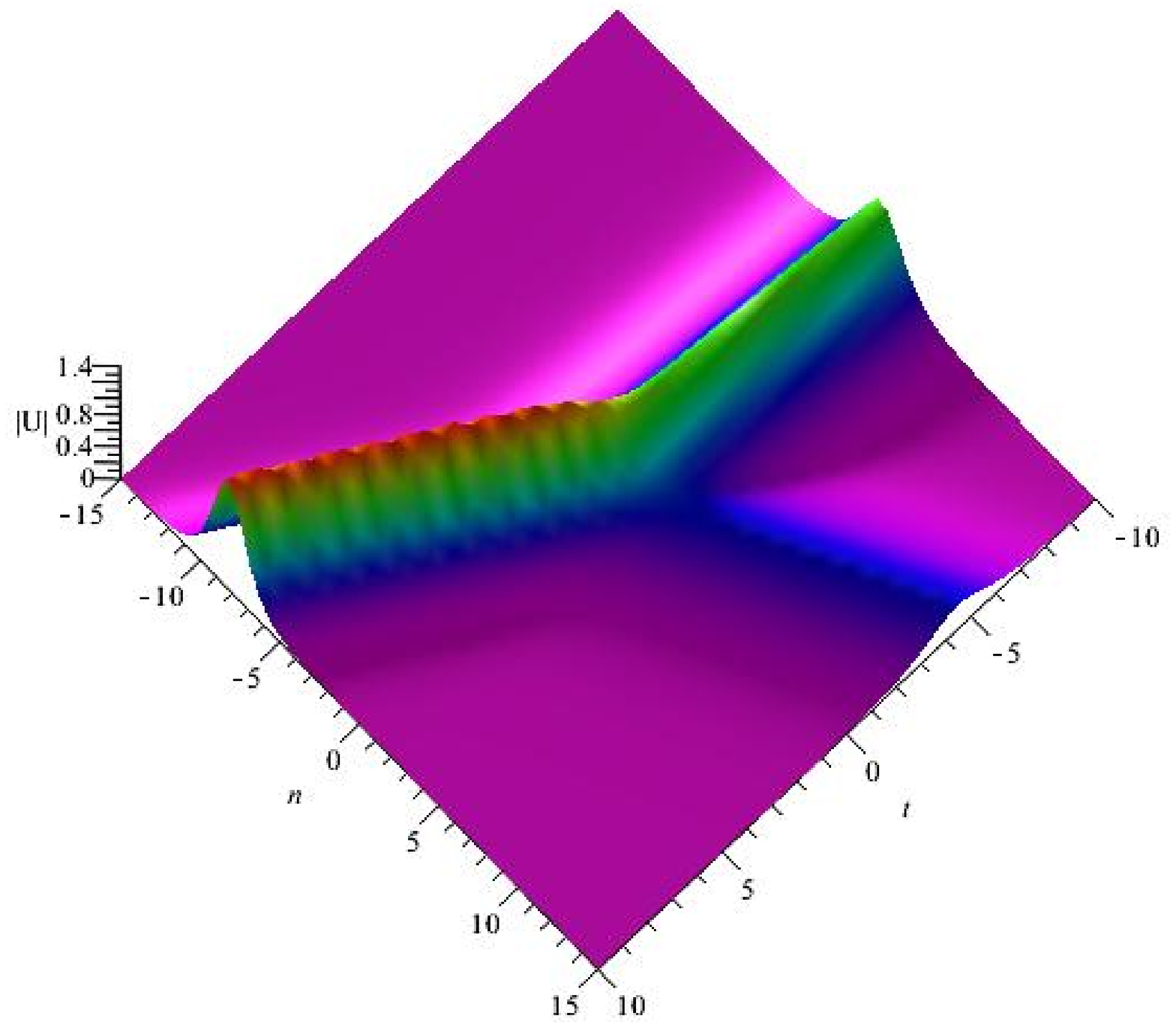}\\
\par(a) fisson & (b) fusion
\end{tabular}\end{center}
\caption{The resonant interaction of two
solitons of discrete RNLS. One soliton splits into two solitons, or, two solitons fuse into one. \label{r-interaction}}
\end{figure}

The coefficient $\chi$ plays a role of classification of soliton
interactions. When $\chi$ is a finite number and not equal to zero,
the regular soliton interaction exists. The resonant
interactions of two solitons occur in Eq. \eqref{semirnls} in the
case of $\chi\rightarrow \infty $ or $\chi\rightarrow 0$. Our
discussion is focused on the case of $\chi = 0$ since the other case
can be analyzed in a similar way. Parametric conditions is given as
follows
\begin{align}
(e^{\beta'_1}-e^{\beta'_2})(e^{\beta_1}-e^{\beta_2})=0.\label{factor}
\end{align}
Choosing different factors of \eqref{factor} as zero, we get two
types of soliton resonances, i.e., the fission and fusion. Fig.\ref{r-interaction} (a)
describes one soliton breaks into two solitons with
the parameters selected as
$\beta_1=0.9,\beta_2=1.8,\beta'_1=0.3,\beta'_2=0.3$. Fig.\ref{r-interaction} (b) shows two solitons fuse into one soliton with the
parameters $\beta_1=0.8, \beta_2=0.8,\beta'_1=1.5,\beta'_2=0.9$.

Similar as the two-soliton solution \eqref{2st1}-\eqref{2st2} for
the continuous equation, we can use the following compact expression
for the above two-soliton solution,
\begin{align}
&f_n=1+a(1,1^*)\exp(\eta_1+\eta'_1)+a(1,2^*)\exp(\eta_1+\eta'_2)\nonumber\\
&\qquad+a(2,1^*)\exp(\eta_2+\eta'_1)+a(2,2^*)\exp(\eta_2+\eta'_2)\nonumber\\
&\qquad+a(1,2,1^*,2^*)\exp(\eta_1+\eta_2+\eta'_1+\eta'_2),\\
&g_n=\exp(\eta_1)+\exp(\eta_2)+a(1,2,1^*)\exp(\eta_1+\eta_2+\eta'_1)\nonumber\\
&\qquad+a(1,2,2^*)\exp(\eta_1+\eta_2+\eta'_2),\nonumber\\
&h_n=\exp(\eta'_1)+\exp(\eta'_2)+a(1,1^*,2^*)\exp(\eta_1+\eta'_1+\eta'_2)\nonumber\\
&\qquad+a(2,1^*,2^*)\exp(\eta_2+\eta'_1+\eta'_2),
\end{align}
where the coefficients are defined by
\begin{align}
& a(i,j)=4\sinh^2(\frac{\beta_i-\beta_j}{2}),\label{re1}\\
& a(i,j^*)=\frac{1}{4}\csch^2(\frac{\beta_i+\beta'_j}{2}),\\
& a(i^*,j^*)=4\sinh^2(\frac{\beta'_i-\beta'_j}{2}).
\end{align}

The exact $N$-soliton solution of
eqs. \eqref{sembb1}-\eqref{sembb3} is presented in the form
\begin{eqnarray}
&& f_n=\sum_{\mu=0,1}^{(e)}\exp\left[\sum_{j=1}^{N}\mu_j\eta_j
+\sum_{j=N+1}^{2N}\mu_j\eta'_{j-N}+\sum_{1\leq
k<l}^{2N}\mu_k\mu_lA_{kl} \right],\label{solu1}\\
&& g_n=\sum_{\nu=0,1}^{(o)}\exp\left[\sum_{j=1}^{N}\nu_j\eta_j+
\sum_{j=N+1}^{2N}\nu_j\eta'_{j-N}+\sum_{1\leq
k<l}^{2N}\nu_k\nu_lA_{kl}
\right],\label{solu2}\\
&&
h_n=\sum_{\lambda=0,1}^{(o)}\exp\left[\sum_{j=1}^{N}\lambda_j\eta_j+
\sum_{j=N+1}^{2N}\lambda_j\eta'_{j-N}+\sum_{1\leq
k<l}^{2N}\lambda_k\lambda_lA_{kl} \right],\label{solu3}
\end{eqnarray}
where
\begin{eqnarray}
&& \eta_j=\alpha_j t+\beta_j n+\gamma_j,\quad
\alpha_j=-4\sinh^2\frac{\beta_j}{2},\quad
j=1,2,\cdots,N,\\
&& \eta'_j=\alpha'_j t+\beta'_j n+\gamma'_j,\quad
\alpha'_j=4\sinh^2\frac{\beta'_j}{2},\quad
j=1,2,\cdots,N,\\
&& \exp(A_{kl})=4\sinh^2(\frac{\beta_k-\beta_l}{2}),\quad k<l=2,3,\cdots, N,\\
&&
\exp(A_{k,N+l})=\frac{1}{4}\csch^2(\frac{\beta_k+\beta'_l}{2}),\quad
k,l=1,2,\cdots,N,\\
&& \exp(A_{N+k,N+l})=4\sinh^2(\frac{\beta'_k-\beta'_l}{2}),\quad
k<l=2,3,\cdots,N.
\end{eqnarray}
The
summations $\sum_{\mu=0,1}^{(e)},\sum_{\nu=0,1}^{(o)}$ and
$\sum_{\lambda=0,1}^{(o)}$ satisfy the condition
\eqref{mmu1}-\eqref{mmu3} respectively.

\section{Conclusions}
To summarize, we presented here one semi-discrete integrable
version  for the (2+1)-dimensional modified HF equation and one semi-discrete version for
the resonant NLS equation. $N-$soliton solution to the both discrete systems are given in standard
form. Interaction properties of two-soliton solutions are analyzed. The regular,
intermediate-state and resonant soliton interactions of the
semi-discrete systems are discussed. Those analysis on the
resonant interactions might have the applications in optical
communication systems. Fully discrete integrable version for the two equations is under
consideration.

\section*{\bf Acknowledgements}
The work was supported in part by the Research Grants Council
of the Hong Kong Special Administrative Region, China (GRF600806).
and National Natural Science Foundation of China (Grant no.11371251). 

\end{document}